\documentclass[10pt,conference]{IEEEtran}
\usepackage{cite}
\usepackage{amsmath,amssymb,amsfonts}
\usepackage{algorithmic}
\usepackage{graphicx}
\usepackage{textcomp}
\usepackage{xcolor}
\usepackage[hyphens]{url}
\usepackage{fancyhdr}
\usepackage{hyperref}

\usepackage{multirow}
\usepackage{comment}
\usepackage{xspace}

\newcommand{\rfloat}{\textsc{ReFloat}\xspace}
\newcommand{\lhs}[1]{\textcolor{blue}{#1}}

\usepackage{newfloat}
\DeclareFloatingEnvironment[
    fileext=los,
    listname={List of Code},
    name=Code,
    placement=tbhp,
]{mycode}

\usepackage{listings}
\lstdefinestyle{MyScala}{
  numbers=left,  
  firstnumber=1,
  numberfirstline=true,
  numberstyle=\scriptsize\ttfamily,
  language=scala,
  aboveskip=3mm,
  belowskip=3mm,
  showstringspaces=false,
  basicstyle={\scriptsize\ttfamily},
  keywordstyle=\bfseries,
  captionpos=b,
  columns=flexible,
  xleftmargin=0.05\textwidth, xrightmargin=0\textwidth,
  breaklines=true,
  breakatwhitespace=true,
  mathescape=true,
  tabsize=4
}

\usepackage{tikz}
\newcommand{\ballnumber}[1]{\tikz[baseline=(myanchor.base)] \node[circle,fill=.,inner sep=1pt] (myanchor) {\color{-.}\bfseries\footnotesize #1};}

\title{\rfloat: 
Low-Cost Floating-Point Processing in ReRAM for Accelerating Iterative Linear Solvers
}

\def\hpcacameraready{} 
\newcommand\hpcaauthors{
Linghao Song$\dagger$, 
Fan Chen$\ddagger$,
Xuehai Qian$\$$,
Hai Li$\star$, and Yiran Chen$\star$}
\newcommand\hpcaaffiliation{University of California Los Angeles$\dagger$, Indiana University Bloomington$\ddagger$,
Purdue University$\$$,
Duke University$\star$}
\newcommand\hpcaemail{linghaosong@cs.ucla.edu,
fc7@iu.edu,
qian214@purdue.edu,
hai.li@duke.edu,
yiran.chen@duke.edu}

\author{
  \ifdefined\hpcacameraready
    \IEEEauthorblockN{\hpcaauthors{}}
      \IEEEauthorblockA{
        \hpcaaffiliation{} \\
        \hpcaemail{}
      }
  \else
    \IEEEauthorblockN{\normalsize{HPCA \hpcayear{} Submission
      \textbf{\#\hpcasubmissionnumber{}}} \\
      \IEEEauthorblockA{
        Confidential Draft \\
        Do NOT Distribute!!
      }
    }
  \fi 
}

\fancypagestyle{camerareadyfirstpage}{%
  \fancyhead{}
  
  \fancyhead[C]{
    \ifdefined\aeopen
    \parbox[][12mm][t]{13.5cm}{\hpcayear{} IEEE International Symposium on High-Performance Computer Architecture (HPCA)}    
    \else
      \ifdefined\aereviewed
      \parbox[][12mm][t]{13.5cm}{\hpcayear{} IEEE International Symposium on High-Performance Computer Architecture (HPCA)}
      \else
      \ifdefined\aereproduced
      \parbox[][12mm][t]{13.5cm}{\hpcayear{} IEEE International Symposium on High-Performance Computer Architecture (HPCA)}
      \else
      \parbox[][0mm][t]{13.5cm}{\hpcayear{} IEEE International Symposium on High-Performance Computer Architecture (HPCA)}
    \fi 
    \fi 
    \fi 
    \ifdefined\aeopen 
      \includegraphics[width=12mm,height=12mm]{ae-badges/open-research-objects.pdf}
    \fi 
    \ifdefined\aereviewed
      \includegraphics[width=12mm,height=12mm]{ae-badges/research-objects-reviewed.pdf}
    \fi 
    \ifdefined\aereproduced
      \includegraphics[width=12mm,height=12mm]{ae-badges/results-reproduced.pdf}
    \fi
  }
  \fancyfoot[C]{}
}

\begin{document}
\maketitle

\ifdefined\hpcacameraready 
  \pagestyle{empty}
\else
  \thispagestyle{plain}
  \pagestyle{plain}
\fi

\newcommand{\hpcaheight}{0mm}
\ifdefined\eaopen
\renewcommand{\hpcaheight}{12mm}
\fi

\begin{abstract}

Resistive random access memory (ReRAM) is a promising technology that can perform low-cost and in-situ matrix-vector multiplication (MVM) in analog domain. Scientific computing requires high-precision floating-point (FP) processing. However, performing floating-point computation in ReRAM is challenging because of high hardware cost and execution time due to the large FP value range. In this work we present \rfloat, a data format and an accelerator architecture, for low-cost and high-performance floating-point processing in ReRAM for iterative linear solvers. \rfloat matches the ReRAM crossbar hardware and represents a block of FP values with reduced bits and an optimized exponent base for a high range of dynamic representation. Thus, \rfloat achieves less ReRAM crossbar consumption and fewer processing cycles and overcomes the noncovergence issue in a prior work. 
The evaluation on the SuiteSparse matrices shows 
\rfloat achieves $5.02\times$ to $84.28\times$ improvement in terms of solver time compared to a state-of-the-art ReRAM based accelerator.

\end{abstract}

\section{Introduction}
\label{sec:intro}

With the diminishing gain of Moore's Law~\cite{waldrop2016chips} and the end of Dennard scaling~\cite{frank2001device},
general-purpose computing platforms
such as CPUs and GPUs
will no longer benefit from 
shrinking transistor size or 
integrating more cores~\cite{hadi2011darksilicon}.
Thus, domain-specific architectures are critical for
improving the performance and energy efficiency of various applications.
Rather than relying on conventional CMOS technology, 
the emerging non-volatile memory technology such as 
resistive random access memory (ReRAM) is considered as a promising candidate for
implementing processing-in-memory (PIM) accelerators
\cite{bojnordi2016memristive,chi2016prime,shafiee2016isaac, song2017pipelayer,song2018graphr,feinberg2018enabling,imani2019floatpim,hu2016dot,yang2019sparse,ji2019fpsa,ankit2019puma,li2018reram} 
that can provide orders of magnitude improvement of computing efficiency.
Specifically, ReRAM can store data and perform in-situ matrix-vector multiplication (MVM) operations in the analog domain.
Most current ReRAM-based accelerators 
focus on machine learning applications, 
which can accept a low precision, e.g., less than 16-bit fixed-point, 
thanks to the quantization in deep learning \cite{gupta2015deep,courbariaux2015binaryconnect,hubara2017quantized,li2016ternary,han2015deep}.

\begin{figure}[tb]
\centering
\includegraphics[width=0.995\columnwidth]{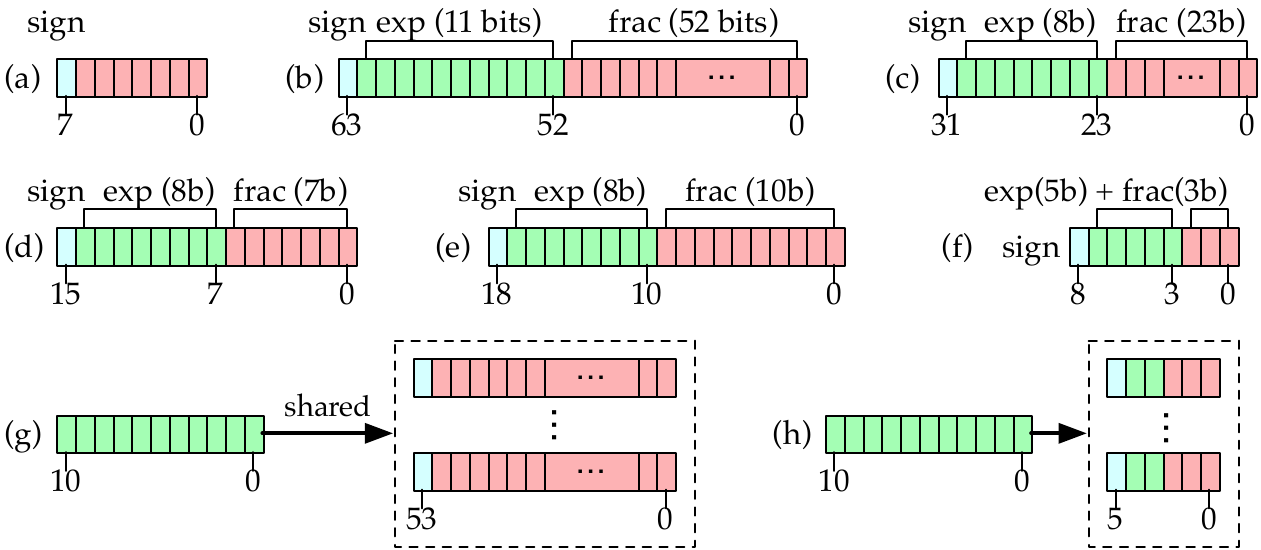}
\vspace{-12pt}
\caption{\lhs{}The bit layout of 
(a) an 8-bit signed integer, 
(b) a 64-bit double-precision floating-point number,
(c) a 32-bit single-precision floating-point number,
(d) a Google bfloat16 number,
(e) an Nvidia TensorFloat32 number,
(f) a Microsoft ms-fp9 number,
(g) a block of numbers in block floating point, and
(h) a block of numbers in {\tt ReFloat}(x,2,3).}
\label{figure:bitlayout}
\vspace{-15pt}
\end{figure}

Scientific computing is a collection of tools, techniques,
and theories for solving science and engineering problems 
modeled in mathematical systems~\cite{golub2014scientific}.
The underlying variables in scientific computing 
are continuous in nature, 
such as time, 
temperature, 
distance, and
density. 
One of the essential aspects of scientific computing 
is modeling a complex system with partial differential equations (PDEs) 
to understand the natural phenomena in science
\cite{harrison2016quantum, jensen2017introduction}, 
or the design and decision-making of engineered systems
\cite{chapman2017accelerated, nobile2017graphics}.
Most problems in continuous mathematics modeled by PDEs
cannot be solved directly.
In practice, the PDEs are converted to a linear system $A\mathbf{x}=\mathbf{b}$,
and then solved through an iterative solver 
that ultimately converges to a numerical solution~\cite{arioli1989solving,saad2003iterative}.
To obtain an acceptable answer 
where the residual is less than
a desired threshold, 
intensive computing power~\cite{fan2004GPU-HPC, song2012multicore-matrix} 
is required to perform the floating-point 
sparse matrix-vector multiplication (SpMV), the
critical computation kernel.

Because of the prevalent floating-point operations in scientific computing, it is desirable to leverage ReRAM to achieve parallel in-situ floating-point SpMV.
When using the ReRAM crossbar to perform SpMV, we partition the matrix into blocks, encode each matrix element as the ReRAM cell conductance, and convert the input vector to wordline voltage through Digital-to-Analog Converters (DACs). 
Thus, the bitline will output the results of the dot-product
between the 
current input vector bits and matrix elements mapped in the 
same crossbar column. 
Each bitline in the output is connected to a sample and hold (S/H) unit.
After all input bits are processed, the results
of the SpMV are available at the output of S/H unit, 
which is converted to multi-bit digital values by 
Analog-to-Digital Converters (ADCs).
In general, the number of bits in the input vector and the matrix determine the number of cycles for performing an SpMV. In contrast, the number of bits representing matrix elements determines the number of crossbars.

We examine mapping the floating-point SpMV 
by leveraging the same principle used in MVM.
Take 64-bit double-precision number as an example:
each floating-point number consists of 
a 1-bit sign ($s$), 
an 11-bit exponent ($e$), 
and a 52-bit fraction ($f$).
The value is interpreted 
as $(-1)^s\times (1.b_{51}b_{50}...b_{0})\times 2^{(e-1023)}$,
yielding a dynamic data range from $\pm 2.2\times 10^{-308}$ to $\pm 1.8 \times 10^{308}$.
The number of crossbars for a matrix $M$ 
increases {\em exponentially}
with the bits number of the exponent ($e_M$) 
and linearly with the bits number of the fraction ($f_M$).
Thus, directly representing floating-point values with a large number of crossbars incurs prohibitive costs.

To reduce the overhead, Feinberg {\em et al.}\cite{feinberg2018enabling} propose to 
truncate the higher bits in exponents,
e.g., using the low 6 bits or module 64 of 
the exponent (the 64 paddings in \cite{feinberg2018enabling})
to represent each original value, 
while keeping the number of fraction
bits the unchanged (52 bits). 
However,
this ad-hoc solution {\em does not ensure the convergence} 
of iterative solves (see Table \ref{tab:crystm03_ites} and Section \ref{sec:perf_cg_bicg}). 
In general, to ensure convergence, we need two requirements. 
(1) correct matrix values, which are 
ensured by \cite{feinberg2018enabling} with the aid of floating-point units (FPUs) 
when the exponent range of a submatrix falls outside the 6 bits mapped to ReRAMs. (2) correct vector values, which is not considered by 
\cite{feinberg2018enabling}. In the computation, matrix value does not change, but vector values change every iteration. Thus, vector values in \cite{feinberg2018enabling} fall out of range (i.e., the 64 padding). 
As a result, the solvers do not converge.
In addition, the hardware cost increases exponentially with the
exponent bits. \cite{feinberg2018enabling} used 6 bits for the exponent, however, we can further reduce the exponent bits. Thus,
\cite{feinberg2018enabling} did not
fully reduce the overhead.

We propose \rfloat, a principled 
approach based on a flexible and fine-grained floating-point
number representation. 
The key insight of our solution is the {\em exponent value locality}
among the elements in a matrix block, which is the granularity
of computation in ReRAM. 
If we consider the whole matrix, the exponent values
can span a wide range, e.g., up to 11 for a matrix, but
the range within a block is smaller, e.g., at most 7
for the same matrix.
It naturally motivates the idea of choosing an {\em exponent base} $e_b$
for all exponents in a block and storing only the 
{\em offsets} from $e_b$. 
For a matrix block, although the absolute exponent values may be large, 
the variation is not.
For most blocks, by choosing a proper $e_b$, the offset values
are much smaller than the absolute exponent values, thereby reducing the 
number of bits required.

Instead of simply using the offset as a lossless compression method, \rfloat aggressively uses fewer bits for exponent offsets ($e$) than the required number of bits to represent them.
The error is bounded by the existence of value locality in real-world matrices. Moreover, the error is refined due to the nature of the iterative solver. Starting from an all-zero vector, an increasingly accurate solution is produced in each iteration. The iterative solver stops when the defined convergence criteria are satisfied.
Because the vector from each iteration is not accurate 
anyway, the computation has certain resilience
to the inaccuracy due to floating-point data representation. 
It is why ~\cite{feinberg2018enabling} can work in certain cases. 
In \rfloat, when an offset is larger (smaller) than the largest (smallest)
offset represented by $e$ bits, the largest (smallest)
value of $e$ bits is used for the offset. 
With $e$-bit exponent offset, the range of exponent values
is $[e_b-2^{(e-1)}+1,e_b+2^{(e-1)}-1]$.
Selecting $e_b$ becomes an optimization problem
that minimizes the difference between the exponents of the original matrix block and the exponents with $e_b$ and $e$-bit offsets.

To facilitate the proposed ideas in a concrete architecture, 
we define the \rfloat format
as {\tt ReFloat}$(b,e,f)(e_v,f_v)$,
where $b$ denotes the matrix block size---the length
and width of a square matrix block is $2^b$,
$e$ and $f$ respectively denote 
the exponent and fraction
bit numbers 
for the matrix, and $(e_v,f_v)$
denotes the exponent and fraction bit numbers for the vector.
With $e_b$ for each block,
we are able to represent
all matrix elements in the block.
Then, we develop the conversion scheme from default 
double-precision floating-point format to \rfloat format
and the computation procedure. 
Based on \rfloat format, we design the low-cost high-performance
floating-point processing architecture in ReRAM.
Our results show that for 12 matrices evaluated
on iterative solvers (CG and BiCGSTAB), only 3 bits for exponent and 8 or 16 bits
for fraction are sufficient to ensure convergence. 
In comparison, \cite{feinberg2018enabling} uses 6 bits for exponent and 
51 bits for fraction without guaranteeing convergence. 
It translates to 
a speedup
of 
$5.02\times$ to $84.28\times$
compared with a state-of-the-art ReRAM-based accelerator \cite{feinberg2018enabling} for scientific computing even with the assumption that the accelerator~\cite{feinberg2018enabling} functions the same as FP64 solvers.
We released the source code at \url{https://github.com/linghaosong/ReFloat}.

\section{Background}
\label{sec:background}

\subsection{In-situ MVM Acceleration in ReRAM}

ReRAM
\cite{wong2012metal, akinaga2010resistive}
has recently demonstrated tremendous potential 
to efficiently accelerate the computing kernels in machine learning.
Conceptually, each element in a matrix $M$
is mapped to the conductance state of a ReRAM cell. At the same time,
the input vector $\mathbf{x}$ is encoded as
voltage levels that are applied on the wordlines of the ReRAM crossbar.
In this way, the current accumulation on bitlines
is proportional to the dot-product of the stored conductance and voltages on the wordlines,
representing the result $\mathbf{y}=M\times \mathbf{x}$.
Such \textit{in-situ} computation significantly reduces the 
expensive memory access in MVM processing 
engines~\cite{hu2016dot},
and most importantly, provide massive opportunities to 
exploit the inherent parallelism in an $N\times{N}$ ReRAM crossbar.

ReRAM-based MVM processing engines are fixed-point hardware in nature 
since the matrix and the vector are respectively represented in {\it discrete} conductance states and voltage levels~\cite{wong2012metal}.
If ReRAM is used to support floating-point MVM operation,
many crossbars will be provisioned for fraction alignment, 
resulting in very high hardware costs.
We will illustrate the problem in Section~\ref{sec:idea} to 
motivate \rfloat design.
Nevertheless, the fixed-point precision requirement is acceptable for machine learning applications thanks to the low-precision and quantized neural network algorithms\cite{kim2015compression,jacob2018quantization,hubara2017quantized,courbariaux2016binarized,gupta2015deep,zhou2017incremental}.
Many fixed-point based accelerators~\cite{bojnordi2016memristive,chi2016prime,shafiee2016isaac,song2017pipelayer,fujiki2018memory,ankit2019puma,ji2019fpsa,yang2019sparse} are built with the ReRAM MVM processing engine and achieve reasonable classification accuracy.

\subsection{Iterative Linear Solvers}

 \begin{mycode}[h]
    \centering
	\vspace{-6pt}
\begin{lstlisting}[style=MyScala]
initiate x = x0
while (not converge) do
	//Step 1: compute the residual
	r = b - A * x
	//Step 2: compute the correction
	compute p 
	//Step 3: update the current solution
	x = x + p
end while
\end{lstlisting}
	\vspace{-9pt}
	\caption{The iterative linear solver.}
	\vspace{-6pt}
	\label{code:ite_solver}
\end{mycode}

Scientific computing is an interdisciplinary science that solves computational problems 
in a wide range of disciplines,
including
physics, mathematics, chemistry, biology, engineering, and other natural sciences
subjects \cite{ferziger2002computational,griewank2008evaluating,antoulas2005approximation}. 
Systems of large-scale PDEs typically model those complex computing problems.
Since it is almost impossible to obtain the analytical solution of those PDEs directly, 
a common practice is to discretize continuous PDEs into a linear system
$A\mathbf{x}=\mathbf{b}$~\cite{arioli1989solving,saad2003iterative}
to be solved by numerical methods.
The numerical solution of this linear system 
is usually obtained by an iterative solver~\cite{wilkinson1994rounding,moler1967iterative,demmel1997applied}.

Code \ref{code:ite_solver} illustrates a typical computing process in iterative methods.
The vector $\mathbf{x}$ to be solved is 
typically initialized to 
an all-zero vector $\mathbf{x}_0$,
followed by three steps in the main body:
(1) the residual (error) of the current solution vector is calculated as 
$\mathbf{r} = \mathbf{b} - A\mathbf{x}$;
(2) to improve the performance of the estimated solution,
a correction vector $\mathbf{p}$ is computed based on the current residual $\mathbf{r}$; and
(3) the current solution vector is improved by adding the correction vector as
$\mathbf{x} = \mathbf{x} + \mathbf{p}$,
aiming to reduce the possible residuals produced in the next calculation iteration.
The iterative solver stops when the defined convergence criteria are satisfied.
Two widely used convergence criteria are
(1) that the iteration index is less than a preset threshold $K$, or (2) that the L-2 norm of the residual  ($\text{res}=||\mathbf{b} - A\mathbf{x}||^2$) is less than a preset threshold $\tau$.
Notably, all the values involved in Code \ref{code:ite_solver}
are implemented as double-precision floating-point numbers
to meet the high-precision requirement of mainstream scientific applications.

The various iterative methods follow the above computational steps
and differ only in calculating
the correction vectors.
Among all candidate solutions, 
Krylov subspace approach is the standard method nowadays.
In this paper, 
we focus on two representative Krylov subspace solvers -- Conjugate 
Gradient (CG)~\cite{hestenes1952methods} and
Stabilized BiConjugate Gradient (BiCGSTAB)~\cite{van1992bi}.
The computational kernels of these two methods are 
large-scale sparse floating-point matrix-vector multiplication $\mathbf{y} = A\mathbf{x}$,
which requires the support of floating-point computation
in ReRAM and imposes significant challenges to the underlying computing hardware.

\subsection{Fixed-Point and Floating-Point Representations}

We use the 8-bit signed integer and
the IEEE 754-2008 standard~\cite{ieee2008754} 64-bit double-precision floating-point number
as examples to compare the difference between fixed-point and floating-point numbers.
They refer to the format used to store and manipulate the digital representation of data.
As shown in Figure~\ref{figure:bitlayout} (a),
fixed-point numbers represent integers---positive and negative whole numbers---via a sign bit followed by multiple ({\it e.g., i}-bit) value bits, yielding a value range of $-2^i$ to $2^i-1$.
IEEE 754 double-precision floating-point numbers shown in Figure~\ref{figure:bitlayout} (b) are designed to represent and manipulate rational numbers,
where a number is represented with a sign bit ($s$), 
an 11-bit exponent ($e$),
and a 52-bit fraction ($b_{51}b_{50}...b_{0}$).
The value of
a double-precision floating-point is interpreted 
as $(-1)^s\times (1.b_{51}b_{50}...b_{0})\times 2^{(e-1023)}$,
yielding a dynamic data range from $\pm 2.2\times 10^{-308}$ to $\pm 1.8 \times 10^{308}$.

Many efficient floating point formats shown in Figure~\ref{figure:bitlayout} have been proposed 
because the default format incurs a
high cost for conventional digital systems.
However, the applications such as
deep learning
do not require a very wide data range. 
The representative examples 
include IEEE 32-bit single-precision floating point (FP32), 
Google bfloat16 \cite{wang2019bfloat16},
Nvidia TensorFloat32 \cite{nvidiaTF32},
Microsoft ms-fp9~\footnote{We infer the layout from the description in \cite{chung2018serving}. No public specifications on ms-fp are available.}~\cite{chung2018serving},
and block floating point (BFP) \cite{bower1990nicam,klank1989dsr}.
Accordingly, 
specialized hardware designs or/and 
systems are also proposed to amplify
the benefits 
of efficient data formats.
For example, Google bfloat is associated with TPU \cite{jouppi2017datacenter,tpu,TPU-2}, 
Nvidia TensorFloat is associated with tensor core GPUs, 
Microsoft floating-point formats are associated with Project Brainwave \cite{chung2018serving}, and
BFP are favorable for signal processing on
DSPs \cite{elam2003block} and FPGAs~\cite{altera2005bfp}.

However, the floating-point 
representations 
favored by deep learning
may not benefit scientific computing.
For deep learning, 
weights can be retrained 
to a narrowed/shrunk space, even without 
floating-point \cite{hubara2017quantized,zhou2016dorefa,lin2015neural,rastegari2016xnor,courbariaux2015binaryconnect}.
In scientific 
computing, data cannot be retrained, and the
shrunk formats can not capture all values.
For example, $1.0\times10^{-40}$ falls out of
range for FP32, bfloat16, TensorFloat32, 
and ms-fp9 because of narrow range representation.
Two values $1.0\times10^{-40}$ and $1.0\times10^{-30}$ can
not be captured by a BFP block because of non-dynamic range 
representation within a block. The narrow 
or non-dynamic range may
lead to non-convergence in 
scientific computing.

In general, double-precision floating-point is a norm for high-precision scientific computations 
because it can support a wide range of data values
with high precision.
However, the processing demands low hardware costs and high performance.

\section{Motivation and ReFloat Ideas}
\label{sec:idea}

\subsection{Fixed-Point MVM processing in ReRAM}
\label{sec:fixedpoint}

\begin{figure}[tb]
\vspace{0pt}
\centering
\includegraphics[width=0.95\columnwidth]{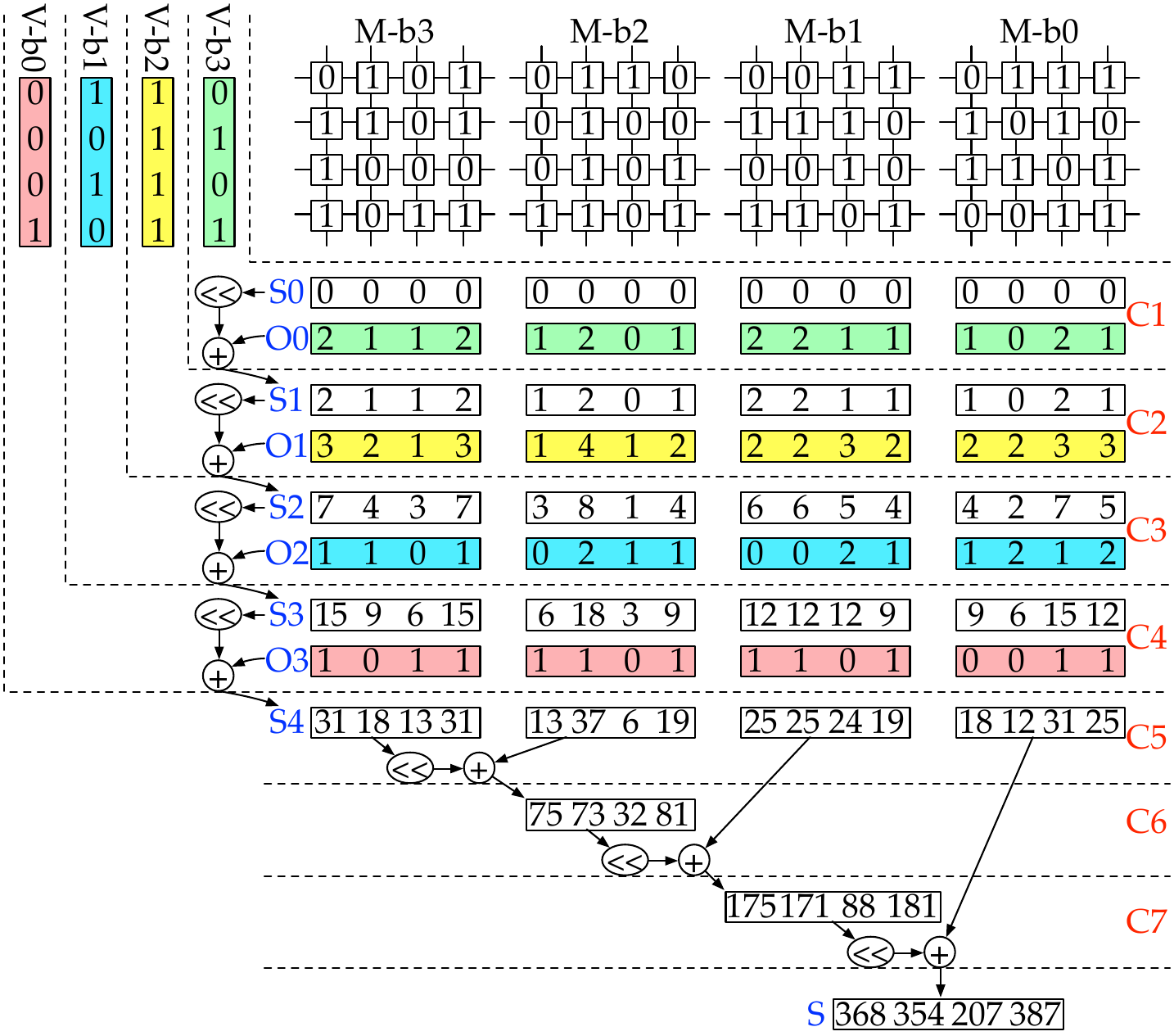}
\vspace{-6pt}
\caption{Fixed-point (integer) MVM in ReRAM.}
\label{figure:fix_point_example}
\vspace{-6pt}
\end{figure}

The processing of SpMV on ReRAM-based accelerators 
utilizes matrix blocking on a large matrix
to perform MVM on matrix blocks
with ReRAM crossbars \cite{feinberg2018enabling,song2018graphr}.
The floating-point MVM is built on fixed-point MVM.
To understand the cycle numbers and ReRAM crossbar numbers in ReRAM-based fixed-point MVM, we use 
Figure~\ref{figure:fix_point_example} as an example.
\begin{equation} 
\label{eq:fix_point_example}
\begin{split}
\left[
\begin{array}{c}
368 \\
354 \\
207 \\
387 \\
\end{array}
\right]_d
&=
\left[
\begin{array}{cccc}
0 & 13 & 7 & 11 \\
11& 14 & 3 & 8 \\
9 & 5 & 2 & 5 \\
14 & 6 & 9 & 15\\
\end{array}
\right]_d^T\times
\left[
\begin{array}{c}
6 \\
12 \\
6 \\
13 \\
\end{array}
\right]_d\\
&= 
\left[
\begin{array}{cccc}
0000 & 1101 & 0111 & 1011 \\
1011 & 1110 & 0011 & 1000 \\
1001 & 0101 & 0010 & 0101 \\
1110 & 0110 & 1001 & 1111\\
\end{array}
\right]_b^T\times
\left[
\begin{array}{c}
0110 \\
1100 \\
0110 \\
1101 \\
\end{array}
\right]_b
\end{split}.
\vspace{-9pt}
\end{equation}
Figure~\ref{figure:fix_point_example}
shows
the processing of fixed-point MVM in ReRAM,
which represents the 
computation of an example Eq.~(\ref{eq:fix_point_example}) by utilizing ReRAM-based MVM engines with single-bit precision.
Before computation, we convert the decimal integers in both the matrix and the vector to binary bits. We set the precision for the matrix and input vector to 4-bit.
The matrix is bit-sliced into 
four 1-bit matrices
and then mapped to four crossbars, i.e., 
M-b3, M-b2, M-b1, and M-b0. 
The input vector is bit-sliced into 
4 one-bit vectors, i.e., 
V-b3, V-b2, V-b1, and V-b0.
The multiplication is performed
in pipeline. 
Each crossbar has a zero initial vector S0.
In the first cycle C1,
the most significant bit (MSB) vector
V-b3 is applied on wordlines of the four
crossbars, and the multiplication results
of V-b3 with M-b3, M-b2, M-b1, and M-b0
are obtained in parallel,
denoted by O0.
In cycle C2, S0 is right-shifted by 1 bit
to get S1,
and V-b2 is input to the crossbars to get
the multiplication results O1. Similar operations
are performed in C3 and C4. After C4, we get S4,
the multiplication results of the input vector 
with four bit-slices of the matrix. 
In the following threes cycles
C5 to C7, we shift and add S4 from the four crossbars
to get the final multiplication result.
For the fixed-point
multiplication of an $N_M$-bit matrix 
with an $N_v$-bit vector, the processing 
cycle count is $C_{\text{int}}=N_v + (N_M - 1)$.

\subsection{Hardware Cost and Performance Analysis of Floating-Point MVM in ReRAM}
\label{sec:motivation}

In this section, we explain in detail the hardware cost, i.e., the {\em crossbar number} $C$,
and the performance, i.e., the {\em cycle number} $T$, of ReRAM-based floating-point MVM.
Note that $C$ correlates with the 
ability to execute floating-point MVMs in parallel 
with a given number of on-chip ReRAMs
\cite{feinberg2018enabling,shafiee2016isaac,song2018graphr}:
the smaller $C$, the more parallelism can be explored.
A smaller $T$ directly reflects a higher performance of
one ReRAM-based MVM on a matrix block. 
A smaller $T$ and a smaller $C$
reflects a higher performance of one SpMV
on a whole matrix.

\vspace{3pt}
\noindent
\textbf{Crossbar number.}
Suppose we compute the multiplication of
a matrix block $M$ and a vector segment $v$.
In the matrix block $M$, the number of fraction bits
is $f_M$ and the number of exponent bits is $e_M$.
In the vector segment $v$, the number of fraction bits
is $f_v$ and the number of exponent bits is $e_v$.
To map the matrix fraction to ReRAM crossbars,
we need $(f_M+1)$ ReRAM crossbars because the fraction is
normalized to a value with a leading 1. For example,
(52+1) crossbars are needed to represent the 52-bit fraction
in double floating-point precision in \cite{feinberg2018enabling}.
To map the matrix exponent to ReRAM crossbars,
we need $2^{e_M}$ ReRAM crossbars for $e_M$-bit exponent
states, which is called padding in \cite{feinberg2018enabling} where 64-bit paddings are
needed for an $e_M=6$. Thus, $C$ is calculated as
\vspace{-3pt}
\begin{equation}
\label{eq:xbar_count}
C
=4\times(2^{e_M}+f_M+1),
\vspace{-3pt}
\end{equation}
where the leading multiplier 4 is contributed from sign bits of
the matrix block and the vector segment.

\vspace{3pt}
\noindent
\textbf{Cycle number.} 
We conservatively suppose the precision of 
digital-analog converters is 1-bit as that in 
\cite{feinberg2018enabling,shafiee2016isaac}.
The number of value states in a vector segment
is $(2^{e_v}+f_v+1)$.
For each input state, we need 
$(2^{e_M}+f_M+1)$ to perform shift-and-add
to reduce the partial results from the ReRAM 
crossbars. To achieve higher computation efficiency, 
a pipelined input and reduce scheme \cite{shafiee2016isaac}
can be used. Thus, $T$ is calculated as
\begin{equation} \label{eq:pe_cycle}
T
=(2^{e_v}+f_v+1)+(2^{e_M}+f_M+1)-1.
\vspace{-3pt}
\end{equation}

\begin{figure}[tb]
\centering
\includegraphics[width=0.95\columnwidth]{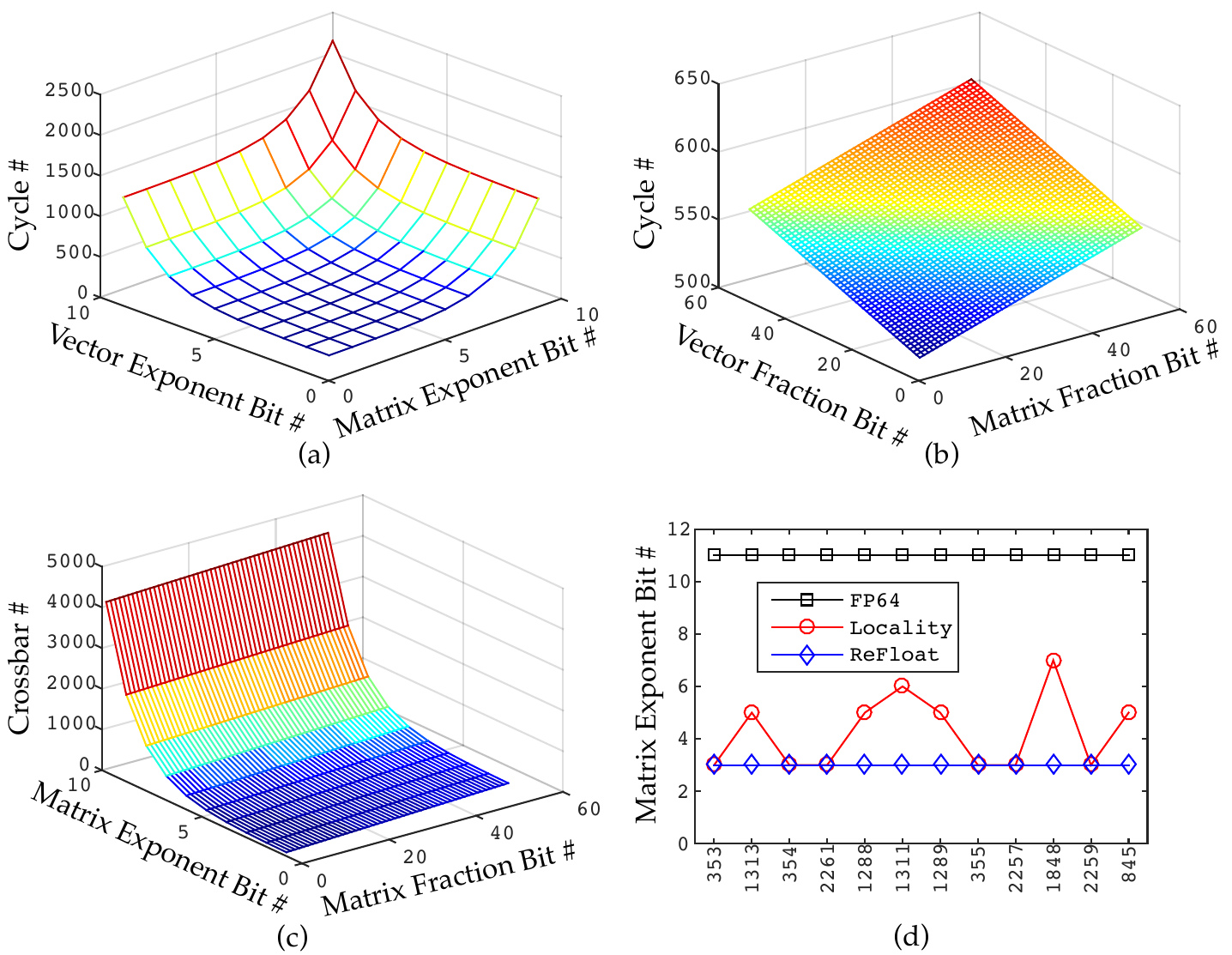}
\vspace{-6pt}
\caption{(a) The cycle number of  various exponent bit number for vector segment and matrix block, (b) the cycle number of various fraction bit number for vector segment and matrix block, (c) the crossbar number of various fraction and fraction bit number for matrix block, and (d) matrix exponent bit number of double-precision floating point(FP64), the locality in 12 matrices, and ReFloat.}
\label{figure:cost_bit_locality}
\vspace{-6pt}
\end{figure}

\vspace{3pt}
\noindent
\textbf{High hardware cost and low performance
in default double precision.} In
double-precision floating-point (FP64),
one MVM in ReRAM consumes
8404 crossbars and 4201 cycles. To understand
how bit number affects the hardware cost 
and performance,
we explore the effect of exponent and
fraction bit number of matrix and vector on
the cycle number and the effect of exponent and
fraction bit number of matrix on the crossbar
number, illustrated in Figure~\ref{figure:cost_bit_locality}.
The crossbar number increases
{\em exponentially} with $e_M$ while linearly with $f_M$.
Furthermore,
the cycle number increases
{\em exponentially} with both $e_v$ and $e_M$,
while the latency increases linearly with $f_v$ and $f_M$.

\subsection{Non-Convergence in \cite{feinberg2018enabling}}

\begin{table}[t]
  \centering
  \caption{The iteration numbers for convergence under various exp(onent) and fra(ction) bit configurations for matrix {\tt crystm03}. \textbf{NC} indicates non-convergence.}
  \vspace{-6pt}

\begin{tabular}{p{4mm}p{8mm}p{8mm}p{8mm}p{8mm}p{8.5mm}p{7.5mm}}
    \hline
\textbf{exp} & \textbf{11} & 11 & 11  & 11 & 11 & 11 \\
\textbf{frac} & \textbf{52} & 30 & 29 & 28 & 27 & 26  \\
\textbf{\#ite} & \textbf{80} & 82(+2) & 82(+2) & 83(+3) & 83(+3) & 84(+4) \\
    \hline
\textbf{exp} & 11 & 11 & 11  & 11 & 11 & 11 \\
\textbf{frac} & 25 & 24 & 23 & 22 & 21 & 20  \\
\textbf{\#ite} & 90(+10) & 93(+13) & 93(+13) & 95(+15) & 107(+27) & \textbf{NC} \\
    \hline
  \end{tabular}
\\ \vspace{2pt}
\begin{tabular}{llllll}
    \hline
\textbf{exp} & 10 & 9  & 8 & 7 & 6 \\
\textbf{frac} & 52 & 52 & 52 & 52 & 52  \\
\textbf{\#ite} & 80 & 80 & 80 & 20620(+256$\times$) & \textbf{NC} \\
    \hline
  \end{tabular}
  \label{tab:crystm03_ites}
\vspace{-3pt}
\end{table}

\label{sec:nc_feinberg}
The above analysis makes it possible to reduce the number of digits by reducing the number of bits of the exponent and fraction, thereby reducing hardware costs, i.e., fewer cycles and crossbars. However, the accuracy of the solvers may be significantly degraded or even cause non-convergence.

The design of the state-of-the-art ReRAM-based accelerator~\cite{feinberg2018enabling} for 
floating-point SpMV
is driven by the goal of 
reducing the number of bits for exponent.
However, this solution adopts an ad-hoc approach that 
simply truncates a number of high order bits in exponent. 
Specifically, with the lower 6 bits of exponent, 
\cite{feinberg2018enabling} uses module 64 of the exponent to represent
each original value and map the matrix to ReRAM. 
For the matrix values out of the range of 6 bits, 
\cite{feinberg2018enabling} uses FPUs to compute.
For the computation of $A\mathbf{x}$, the matrix
$A$ can be accurately processed in \cite{feinberg2018enabling}.
However, the values of vector $\mathbf{x}$ change at every iteration
but \cite{feinberg2018enabling} does not provide any solution for processing correct vector values.
Thus, the vector $\mathbf{x}$ values can fall out of
the ranges of 64 paddings (6 bits), and 
{\em non-convergence} happens in \cite{feinberg2018enabling}.

Table \ref{tab:crystm03_ites} shows the number of iterations for convergence under various exponent and fraction bit configurations.
In default double-precision, it takes 80
iterations to convergence. 
If we fix the exponent bits
and truncate fraction bits, a 21-bit fraction 
takes 27 additional iterations, and a fraction less than
21 bits leads to non-convergence. 
If we fix the fraction bits
and truncate exponent bits like \cite{feinberg2018enabling},
7-bit exponent increases the iteration number from
80 to 20620, and an exponent less than
7 bits leads to non-convergence. 
Thus, the solution proposed in \cite{feinberg2018enabling} 
may break the correctness of the iterative solver. 
In contrast, the number of bits in fraction has less
impact on the number of iterations to converge. 
For example, Table \ref{tab:crystm03_ites} shows that 
drastically reducing fraction bits
from 52 to 30 only increases
the number of iterations by 2$\times$. 
However, ~\cite{feinberg2018enabling} kept the 
number of bits in fraction unchanged and lost the opportunity
to reduce hardware cost and improve performance. 
Thus, we are convinced that we need to develop a more principled approach to find a better solution to the problem.

\subsection{Value Locality \& Bit Compression}

We leverage an intuitive observation of matrix 
element values---{\em exponent value locality}---to reduce the number
of exponents bits while keeping enough accuracy. 
We define the locality as
the maximum number of required bits to cover
the exponent in all matrix blocks
of a large matrix.
We illustrate the locality of matrices
from SuiteSparse
\cite{davis2011university} in 
Figure~\ref{figure:cost_bit_locality}(d).
As discussed before, ReRAM performs MVM at the granularity of 
matrix block, whose size is determined by the size of ReRAM crossbar, e.g., 128 $\times$ 128.
While exponent values of the whole matrix can span a wide range, e.g., up to 11 for a matrix, but
the range is smaller within a block, e.g., at most 7
for the same matrix. 
Therefore, the default locality, i.e., 11, is redundant.
Naturally, it motivates the idea of using 
an {\em exponent base $e_b$} for all exponents
in a block and storing only the {\em offsets} from $e_b$.
For most blocks, by choosing a proper $e_b$, the offset values
are much smaller than the absolute exponent values, thereby reducing the 
number of bits required.

It is important to note that we do not 
simply use the offset as a lossless compression method.
While exponent value locality exists for most of the blocks,
it is possible that for a small number of blocks, the exponent values
are scattered across a wide range.
If we include enough bits for all offsets, the benefits
for the majority of blocks will be diminished.
Moreover, it is not necessary due to the nature of iterative solvers.

We can naturally tune the 
accuracy by the number of bits $e$ allocated for the offsets,
which is less than the number of exponent bits necessary 
to represent the offsets precisely. 
When an offset is larger (smaller) than the largest (smallest)
offset representable by $e$ bits, the largest (smallest)
value of $e$ bits is used accordingly. 
With $e$-bit exponent offset, the range of exponent values
is $[e_b-2^{(e-1)}+1,e_b+2^{(e-1)}-1]$.
Intuitively, given $e$ and $e_b$, this system can precisely
represent the exponent values that fall into a ``window'' around 
$e_b$, while the ``size of the window'' is determined by $2^{(e-1)}$.
Then, selecting $e_b$ becomes an optimization problem
that minimizes the difference between the exponents of the original matrix block and the exponents with $e_b$ and $e$-bit offsets. Thus,
we achieve
a wide data range but a 
low hardware cost simultaneously.

\section{ReFloat Data Format}
\label{sec:format}

\subsection{ReFloat Format}
\label{sec:format41}

\begin{table}[t]
  \centering
  \caption{List of symbols and descriptions.}
  \vspace{-6pt}
  
  \begin{tabular}{rl}
    \hline
    \multicolumn{2}{l}{{\tt ReFloat}$(b,e,f)(e_v,f_v)$ : {\tt ReFloat} format notation. }\\
    \hline
    \hline
    \textbf{Symbol} & \textbf{Description}\\
    \hline
    $2^b$ & The size of a square block.\\
    $e$ & The number of exponent bits for a matrix block. \\
    $f$ & The number of fraction bits for a matrix block. \\
    \hline
    $A$ & A sparse matrix. \\
    $\mathbf{b}$ & The bias vector for a linear system. \\
    $\mathbf{x}$ & The solution vector for a linear system. \\
    $\mathbf{r}$ & The residual vector for a linear system. \\
    $a$ & A scalar of $A$. \\
    $(a)_e$ & The exponent of $a$, $(a)_e\in\{0,1,2,...\}$. \\
    $(a)_f$ & The fraction of $a$, $(a)_f\in(1,2)$. \\
    $A_c$ & A block of the sparse matrix $A$. \\
    $(i,j)$ & The index for the block $A_c$. \\
    $(ii,jj)$ & The index for the scalar $a$ in the block $A_c$. \\
    $(iii,jjj)$ & The index for the scalar $a$ in the matrix $A$. \\
    $e_b$ & The base for exponents of elements in a block.
    \\
    \hline
    $e_{bv}$ & The base for exponents a vector segment. \\
    $e_v$ & The number of exponent bits for a vector segment. \\
    $f_v$ & The number of fraction bits for a vector segment. \\
    \hline
  \end{tabular}
  \label{table:symbol}
  \vspace{-6pt}
\end{table}

\begin{figure}[tb]
\vspace{0pt}
\centering
\includegraphics[width=0.9\columnwidth]{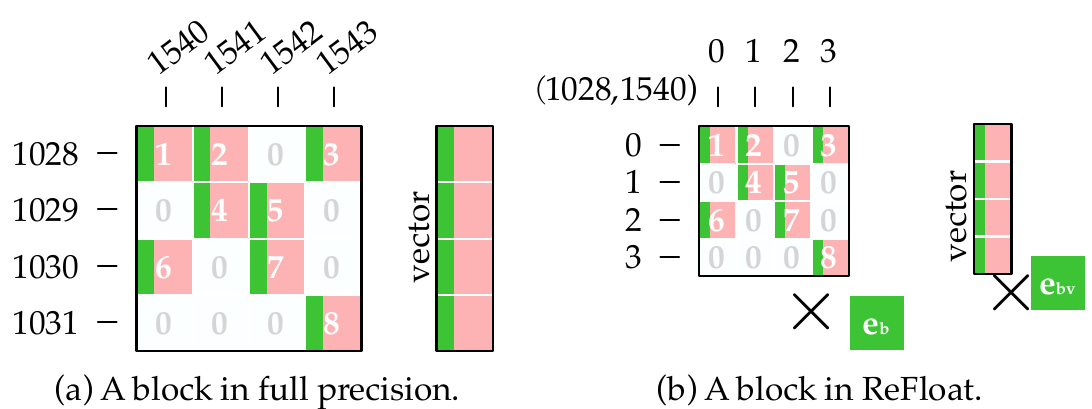}
\vspace{-6pt}
\caption{Comparison of a matrix block (a) in original full precision format and (b) in \rfloat format. }
\label{figure:coo_refloat}
\vspace{-6pt}
\end{figure}

We define
\rfloat format as 
{\tt ReFloat}$(b,e,f)(e_v,f_v)$, 
where $b$ determines the matrix block size $2^b$ (the length
and width of a square matrix block),
$e$ and $f$ respectively denote 
the exponent and fraction
bit numbers 
for the matrix, 
and $(e_v,f_v)$ denotes the bit numbers for the vector.
Table~\ref{table:symbol} lists
the symbols and corresponding descriptions
in \rfloat.

Figure~\ref{figure:coo_refloat} intuitively illustrates
the idea of \rfloat. 
In Figure~\ref{figure:coo_refloat} (a), 
each scalar is in a 64-bit floating-point format.
It requires a 32-bit integer for row index
and a 32-bit integer for column index
to locate each element in the matrix block.
Therefore, we need 
$8\times(32 + 32 + 64)=1024$ bits for storing
the eight scalars. 
With \rfloat, 
assuming we use {\tt ReFloat}$(2,2,3)$ format
as depicted in Figure~\ref{figure:coo_refloat} (b), we see that:
(1) each scalar in the block can be indexed 
by two 2-bit integers;
(2) the element value is represented by a $1+2+3=6$-bit floating point number~\footnote{The elements inside a \rfloat block are floating-point, while the elements inside a BFP block are fixed-point.}; (3) the block is indexed by two 30-bit integers 
and (4) an 11-bit exponent base $e_b$ is also recorded. 
Therefore,
we only use $8\times(2+2+6)+2\times 30+11=151$ bits 
to store the entire matrix block,
which reduces the memory requirement by approximately 10$\times$ (151 vs. 1024). 
This reduction in bit representation is also beneficial for 
reducing the number of ReRAM crossbars for computation 
in hardware implementation.
Specifically, the full precision format consumes 118 crossbars, 
as illustrated in \cite{feinberg2018enabling}, 
our design only requires
16 crossbars with 
{\tt ReFloat}$(2,2,3)$ format.
Thus, given the same chip area, 
our design is able to process
more matrix blocks 
in parallel.

\subsection{Conversion to ReFloat Format}

\begin{figure}[tb]
\centering
\includegraphics[width=0.9\columnwidth]{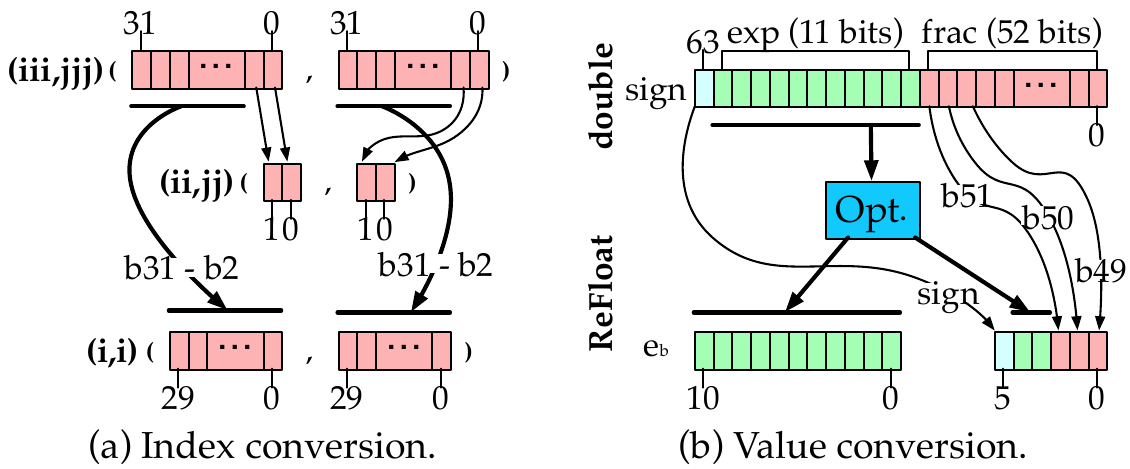}
\vspace{-6pt}
\caption{The conversion of index and value in floating-point format to \rfloat format. }
\label{figure:refloat_conversion}
\vspace{-6pt}
\end{figure}

In order to convert the original matrix to 
a {\tt ReFloat}$(b,e,f)$ format, 
three hyperparameters need to be determined in advance.
The $b$ defines 
how the indices of input data are converted
and determined by the physical size 
of ReRAM crossbars, i.e., a crossbar
with $2^b$ wordlines and $2^b$ bitlines. 
As demonstrated in Figure~\ref{figure:refloat_conversion} (a), 
the leading 30 bits---$b_{31}$ to $b_{2}$ of the index $(iii,jjj)$ 
for a scalar in the matrix $A$---are copied to the same bits in the index
$(i,j)$ for the block $A_c$. 
For each scalar in the block $A_c$, the index 
$(ii,jj)$ for that scalar inside the
block $A_c$ is copied from the last two bits
of the index $(iii,jjj)$. 
The scalars in the same block share
the block index
$(ii,jj)$, and each scalar uses fewer bits for the index inside that block. Thus, we also save
memory space for indices.

\begin{table}[tb]
  \centering
  \caption{Various formats represented by \rfloat.}
  \vspace{-6pt}
  \footnotesize
  \begin{tabular}{rl}
    \hline
Int8 & {\tt ReFloat}$(0,0,7)$ \\
bfloat16 \cite{wang2019bfloat16} & {\tt ReFloat}$(0,8,7)$ \\
Int16 & {\tt ReFloat}$(0,0,15)$ \\
ms-fp9 \cite{chung2018serving} & {\tt ReFloat}$(0,5,3)$ \\
FP32(float) & {\tt ReFloat}$(0,8,23)$ \\
TensorFloat32 \cite{nvidiaTF32} & {\tt ReFloat}$(0,8,10)$ 
\\
FP64(double) & {\tt ReFloat}$(0,11,52)$ \\
BFP64 & {\tt ReFloat}$(6,0,52)$
\\

    \hline
  \end{tabular}
  \label{tab:representation_in_refloat}
\vspace{-6pt}
\end{table}

The hyper-parameters $e$ and $f$ 
determine the accuracy of floating-point values.
A floating-point number consists of three
parts:
(1) the sign bit,
(2) the exponent bits, 
and (3) the fraction bits. 
When converted
to \rfloat, the sign bit remains unchanged.
For the fraction, we only keep the leading
$f$ bits from the original fraction bits
and remove the rest bits in the fraction,
as shown in Figure~\ref{figure:refloat_conversion} (b).
For the exponent bits, we need to first determine the
base value $e_b$ for the exponent. As $e$ means the number
of bits for the ``swing'' range, we need to find an
optimal base value $e_b$ to utilize the $e$ bits fully.
We formalize the problem as an optimization for find the
$e$ to minimize a loss target $L$, defined as 
\vspace{-3pt}
\begin{equation}
\label{eq:loss}
\begin{array}{l}
\min\limits_{e_b} L,   
~L=\sum\limits_{a\in A_c}\left(
\log_2\left(
\frac{a}{(a)_{f}\times 2^{e_b}}
\right)
\right)^{2}
=\sum\limits_{a\in A_c}\left(
(a)_{e}-{e_b}
\right)^{2}.
\end{array}
\vspace{-3pt}
\end{equation}
Let $\partial L/\partial e_b=0$, we can get
\vspace{-6pt}
\begin{equation} \label{eq:eb}
e_b
=\left[
\frac{1}{|A_c|}
\sum_{a\in A_c}(a)_{e}
\right].
\vspace{-3pt}
\end{equation}
Thus, we use the original exponent to 
minus the optimal $e_b$ to get an $e$-bit signed integer
in the conversion.
The $e$-bit signed integer is the exponent in \rfloat.

We use an example to illustrate the format conversion intuitively. The original floating-point
values in Eq. (\ref{eq:v1}) are converted to {\tt ReFloat}(x,2,2)
format in Eq. (\ref{eq:v2}), 
\vspace{-3pt}
\begin{equation} \label{eq:v1}
\left[
\begin{array}{cc}
(-1)\times 1.1111\times 2^7  &  1.0101\times 2^8 \\
(-1)\times 1.0000\times 2^9  &  1.0001\times 2^7
\end{array}
\right]=
\left[
\begin{array}{cc}
-248.0  &  336.0 \\
-512.0  & 136.0
\end{array}
\right],
\end{equation}
\vspace{-6pt}
\begin{equation} \label{eq:v2}
2^8\times\left[
\begin{array}{cc}
(-1)\times 1.11\times 2^{-1}  &  1.01\times 2^0 \\
(-1)\times 1.00\times 2^1  &  1.00\times 2^{-1}
\end{array}
\right]=
\left[
\begin{array}{cc}
-224.0  &  320.0 \\
-512.0  & 128.0
\end{array}
\right],
\vspace{-3pt}
\end{equation}
where $e_b=8$.
Here, we see that \rfloat incurs conversion loss
for the conversion of floating-point values from the original. However,
for scientific computing, the errors
in the iterative solver are gradually corrected. Thus, the errors
introduced by the conversion will also be corrected in the iteration.
From an application/algorithm perspective, \rfloat format is versatile, and the popular formats
in Figure~\ref{figure:bitlayout} can all
be represented by \rfloat as listed in TABLE \ref{tab:representation_in_refloat}. The low hardware cost and format versatility benefit 
the high performance and fast convergence
of \rfloat in solving PDEs. 
We will show the performance and convergence of the iterative 
solver in \rfloat format in Section~\ref{sec:evaluation}.

\subsection{Computation in ReFloat Format}

\begin{figure*}[tb]
\centering
\includegraphics[width=1.95\columnwidth]{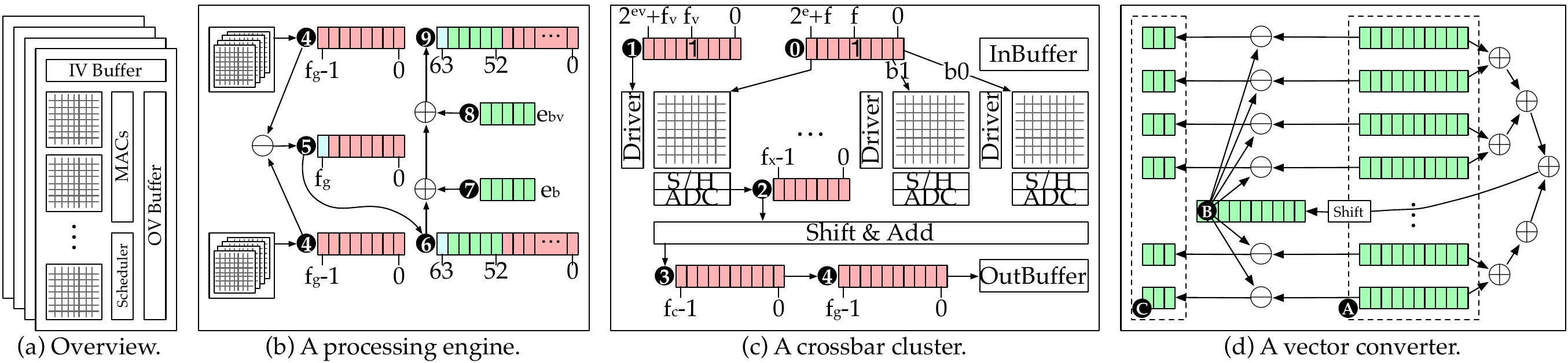}
\vspace{-6pt}
\caption{(a) the accelerator architecture overview. Architectures of (b) a processing engine for floating-point MVM on a matrix block, (c) a crossbar cluster for fixed-point MVM,
and (d) a vector converter. }
\label{figure:processing_engine}
\vspace{-6pt}
\end{figure*}

The matrix $A$ is partitioned into blocks. To compute the 
matrix-vector multiplication $\mathbf{y} = A\mathbf{x}$, the input 
vector $\mathbf{x}$ and the output vector $\mathbf{y}$ are correspondingly
partitioned into vector segments $\mathbf{x}_c$ and $\mathbf{y}_c$. 
The size of the vector segments is $(2^b\times1)$.

For the $p$-th output vector segment $\mathbf{y}_c(p)$, the computation 
in the default full precision will be 
\vspace{-3pt}
\begin{equation} \label{eq:yc}
\mathbf{y}_c(p)
=
\sum\limits_{i}A_c(i,p)\mathbf{x}_c(i),
\vspace{-3pt}
\end{equation}
where $A_c(i,p)$ is the matrix block indexed by $(i,p)$ and $\mathbf{x}_c(i)$
is the input vector segment indexed by $i$. The matrix blocks at the $p$-th 
block column are multiplied with the input vector segments for partial sums
and then they are accumulated.
In the computation for each matrix block,
because the original matrix block $A_c(i,p)$ is converted to
$A_c(i,p)\simeq2^{e_b(i,p)} \tilde{A}_c(i,p)$,
the original vector segment $\mathbf{x}_c(i)$ is converted to
$\mathbf{x}_c(i)\simeq2^{e_{bv}(i)} \tilde{\mathbf{x}}_c(i)$,
and we
encode
$2^{e_b(i,p)} \tilde{A}_c(i,p)$ and  $2^{e_{bv}(i)} \tilde{\mathbf{x}}_c(i)$
by {\tt ReFloat}.
Thus, the multiplication for 
the matrix block $A_c(i,p)$ and the vector segment $\mathbf{x}_c(i)$ is computed as 
$A_c(i,p)\mathbf{x}_c(i)
=
2^{e_b(i,p)+e_{bv}(i)} \tilde{A}_c(i,p)\tilde{\mathbf{x}}_c(i)$.
The matrix-vector multiplication for 
the $p$-th output vector segment in 
the default format Eq. (\ref{eq:yc})
is then computed as 
\vspace{-3pt}
\begin{equation}
\label{eq:yc_new}
\mathbf{y}_c(p)
=
\sum\limits_{i}2^{e_b(i,p)+e_{bv}(i)} \tilde{A}_c(i,p)\tilde{\mathbf{x}}_c(i).
\vspace{-3pt}
\end{equation}
Here we see that
with \rfloat format, the block matrix multiplication
in the default format is preserved.
In the hardware processing,
we perform the fixed-point MVM $\tilde{A}_c\tilde{\mathbf{x}}_c$
by the ReRAM crossbars as shown in Figure~\ref{figure:processing_engine}(c) and multiply the vector exponent
and the block exponent in a processing engine as shown in
Figure~\ref{figure:processing_engine}(b).
Thus, the original high-cost multiplication in full precision
$A_c\mathbf{x}_c$ is replaced by a low-cost multiplication.

\section{ReFloat Accelerator Architecture}
\label{sec:architecture}

\subsection{Accelerator Overview}

Figure~\ref{figure:processing_engine}(a) 
shows the overall architecture of
the proposed accelerator for floating-point scientific computing in ReRAM
with \rfloat. 
We organize 
the accelerator into multiple banks.
Within each bank, ReRAM crossbars are deployed for processing
matrix blocks of floating-point MVM. 
The Input Vector (IV) and 
Output Vector (OV) buffer 
are used for buffering the 
input and output vectors and matrix blocks. 
The Multiply-and-Accumulate (MAC)
units are used to update the vectors.
The scheduler is responsible for the coordination of
the processing.

\subsection{Processing Engine}

The most critical component in the accelerator 
is the processing engine for floating-point
SpMV
in \rfloat format. 
The processing engine consists of a 
few ReRAM crossbars 
and several peripheral functional units.
We show the
architecture of the processing engine in 
Figure~\ref{figure:processing_engine}(b),
assuming we are performing the floating-point SpMV
on a matrix block with the format {\tt ReFloat}$(b,e,f)$.

The inputs to the processing engine are: 
(1) a matrix block 
in {\tt ReFloat}$(b,e,f)$ format; 
(2) an input vector segment in
floating-point with $e_\text{v}$
exponent bits and
$f_\text{v}$ fraction bits 
and the vector length
is $2^b$; 
and (3) the exponent base bits $e_b$ for each matrix block.
The output of a processing engine is a vector segment
for SpMV on the matrix block, 
which is a  double-precision floating-point number.

Before the computation, the matrix block is mapped to the ReRAM
crossbars as detailed in Figure~\ref{figure:processing_engine}(c). 
The fraction part
of the matrix block
in {\tt ReFloat}$(b,e,f)$ represents a number of $1.b_{f-1}...b_0$, then we have $(f+1)$ bits for mapping.
The $e$-bit exponent of the matrix block 
contributes to $2^e$ padding bits
for alignment, then we have another $2^e$ bits for mapping.
Thus, we map the total $(2^e+f+1)$ bits \ballnumber{0} to $(2^e+f+1)$ ReRAM crossbars, where the $i$-th bits of the matrix block is mapped to the $i$-th crossbar~\footnote{Here, we assume that the cell precision for the ReRAM crossbars is 1-bit.
For 2-bit cells, two consecutive bits are mapped to a crossbar.}.
For the input vector segments with $e_v$ exponent bits
and $f_v$ exponent bits, a total number of $(2^{e_v}+f_\text{v}+1)$ bits \ballnumber{1}
are applied to the driver.

During processing, a cluster of crossbars are deployed
to perform the fixed-point MVM for the fraction
part of the input 
vector segment
with the fraction part of the matrix block
using
the shift-and-add method, as the example 
in Figure~\ref{figure:fix_point_example}.
The input bits are applied to the crossbars by
the driver and the output from the crossbar is
buffed by a Sample/Hold (S/H) unit and then converted
to digital by a shared Analog/Digital Converter 
(ADC).
For each input bit to the driver (we assume an 1-bit DAC), 
as the crossbar size is $2^b$, 
the ADC conversion precision is
$f_x=b$ bits.
Then we
need to shift-and-add the results \ballnumber{2} from all $(2^e+f+1)$ crossbars
to get the results \ballnumber{3} for the 
1-bit multiplication
of the vector with the matrix fraction.
Thus, the  bits number of \ballnumber{3} is
$f_c=
2^e+f+1+b$.
Next, we sequentially input the bits in 
\ballnumber{1} to the crossbars and
shift-and-add the collected \ballnumber{3}
for each of the $(2^{e_v}+f_\text{v}+1)$ bits
to get \ballnumber{4}, 
which is the result for the multiplication of the 
matrix block with the input \ballnumber{1}. The bits number
of \ballnumber{4} is
$f_g=
f_c+2^{e_v}+f_\text{v}+1+b$.
As shown in Figure~\ref{figure:processing_engine}(b),
each matrix block has a sign bit,
therefore, 
it requires two crossbar clusters in a processing engine
for the signed multiplication.
Each element in the input vector segment also
has a sign bit. Thus, we need four \ballnumber{4} and 
subtract them to get \ballnumber{5}, 
which is the multiplication results between the matrix block 
and the vector segment. The number of bits for \ballnumber{5}
is $(f_g+1)$, and \ballnumber{5} is a
signed number due to the subtraction.
Next, we convert the \ballnumber{5} to a
double-precision floating-point \ballnumber{6}.
$e_b$ \ballnumber{7} is the 
exponent base for the matrix block
 and
$e_v$ \ballnumber{8} is 
exponent for the vector segment. We 
add \ballnumber{7} and \ballnumber{8} to the exponent of \ballnumber{9}
to get the \ballnumber{9}---
the final results for the multiplication of 
the matrix block with the vector segment in 64-bit double-precision floating-point.

The vector converter is responsible for
converting a vector segment in default floating-point precision to \rfloat
for processing in next iteration.
\ballnumber{A} the exponents of elements in a vector segment is accumulated by an adder tree
and shifted following Eq. (\ref{eq:eb})
to get \ballnumber{B} the vector exponent base $e_{bv}$. An element-wise subtraction is performed on \ballnumber{A} to
get \ballnumber{C} the exponents of the elements in the
vector segment.

\subsection{Streaming and Scheduling}
\label{sec:streaming}

\begin{figure}[tb]
\centering
\includegraphics[width=0.9\columnwidth]{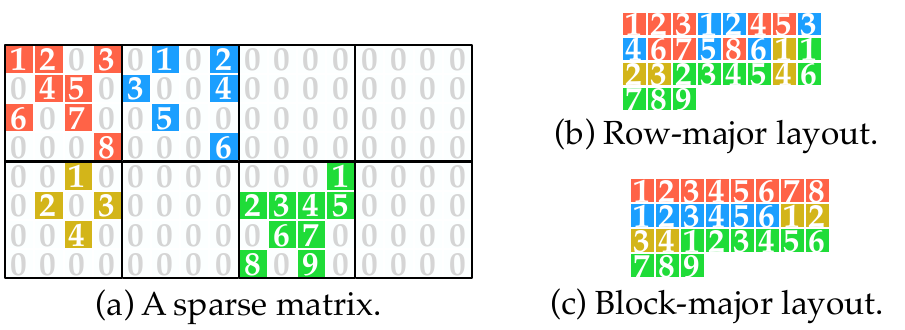}
\vspace{-6pt}
\caption{The row-major layout and block-major layout of a sparse matrix.}
\label{figure:layout}
\vspace{-6pt}
\end{figure}

For the original large-scale  sparse matrix, the non-zero
elements are stored in either row-major or column-major order.
However, the computation in ReRAM crossbars requires 
accessing elements in a matrix block, i.e., elements
indexed by the same window of rows and columns.
Thus, there is a mismatch between the data storage format
in the original application, e.g., Matrix Market File Format~\cite{boisvert1997matrix}, and the most suitable format for \rfloat accelerator.
Direct access to the elements in each matrix block 
will result in random access and wasted memory bandwidth.
We propose a block-major layout to overcome this problem, which ensures that most matrix block elements can be read sequentially.
Specifically, the non-zeros of each $2^b \times 2^b$ block
are stored consecutively, and the non-zeros of every $P$ blocks
among the same set of rows are stored linearly before
moving to a different set of rows, as shown in Figure~\ref{figure:layout}.
Here, $P$ is the number of 
blocks that can be processed in parallel, which is determined by 
the hyper-parameters $b$, $e$, and $f$ for a given number of
available ReRAM crossbars. 

\begin{table}[tb]
  \centering
  \caption{Platform Configuration.}
  \vspace{-6pt}
  \begin{tabular}{ll|ll}
    \hline
    \multicolumn{4}{c}{\textbf{GPU} (Tesla V100 SXM2)}\\
    \hline
    Architecture & Volta & CUDA Cores & 5120\\
    Memory & 32GB HBM2 & CUDA Version & 11.7\\
    \hline
    \multicolumn{4}{c}{\textbf{Feinberg}~\cite{feinberg2018enabling}}\\
    \hline
    Bank & 128 & Crossbar Size & $128\times128$\\
    Clusters/Bank & 64 & Precision & \texttt{\textbf{double}} \\
    Xbars/Cluster & 128 & Comp. ReRAM & 17.1Gb \\
    \hline
    \multicolumn{4}{c}{\textbf{ReFloat}}\\
    \hline
    Bank & 128 & Crossbar Size & $128\times128$\\
    Subbank & 128 & Precision & \texttt{\textbf{refloat}} \\
    Xbars/Subbank & 64 & Comp. ReRAM & 17.1Gb \\
    \hline
    \multicolumn{4}{c}{\textbf{ADC}}\\
    \hline
    \multicolumn{4}{c}{10-bit pipelined SAR ADC @ 1.5GS/s }\\
    \hline
    \multicolumn{4}{c}{\textbf{ReRAM Cells}}\\
    \hline
    \multicolumn{4}{c}{1-bit SLC, $T_{\text{w}}=50.88$ns,
    Comp. Latency=107ns @ (128$\times$128).}\\
    \hline
    
  \end{tabular}
  \label{table:platform}
  \vspace{-6pt}
\end{table}

\section{Evaluation}
\label{sec:evaluation}

\subsection{Evaluation Setup}

\begin{table}[tb]
  \centering
  \vspace{3pt}
\caption{Matrices in the evaluation.}
\vspace{-6pt}
  \begin{tabular}
  {p{5.5mm}p{20mm}p{9mm}p{10.5mm}p{9mm}p{9mm}}
    \hline
    \textbf{ID} & \textbf{Name}& \textbf{\#Rows}& \textbf{NNZ} & \textbf{NNZ/R} & $\kappa$\\
    \hline
353  & crystm01         & 4,875   & 105,339 & 21.6 & 4.21e+2 \\
1313 & minsurfo         & 40,806  & 203,622 & 5.0  & 8.11e+1 \\
354  & crystm02         & 13,965  & 322,905 & 23.1 & 4.49e+2 \\
2261 & shallow\_water1  & 81,920  & 327,680 & 4.0  & 3.63e+0 \\
1288 & wathen100        & 30,401  & 471,601 & 15.5 & 8.24e+3 \\
1311 & gridgena         & 48,962  & 512,084 & 10.5 & 5.74e+5 \\
1289 & wathen120        & 36,441  & 565,761 & 15.5 & 4.05e+3 \\
355  & crystm03         & 24,696  & 583,770 & 23.6 & 4.68e+2 \\
2257 & thermomech\_TC   & 102,158 & 711,558 & 6.9  & 1.23e+2 \\
1848 & Dubcova2         & 65,025  & 1,030,225 & 15.84 & 1.04e+4 \\
2259 & thermomech\_dM   & 204,316 & 1,423,116 & 6.9 & 1.24e+2 \\
845  & qa8fm            & 66,127  & 1,660,579 & 25.1 & 1.10e+2 \\
    \hline
\multicolumn{6}{c}{

\begin{tabular}{c@{\hskip 0.1mm}c@{\hskip 0.1mm}c@{\hskip 0.1mm}c@{\hskip 0.1mm}c@{\hskip 0.1mm}c@{\hskip 0.1mm}c@{\hskip 0.1mm}c@{\hskip 0.1mm}c@{\hskip 0.1mm}c@{\hskip 0.1mm}c@{\hskip 0.1mm}c@{\hskip 0.1mm}}
353 &
1313 &
354 &
2261 &
1288 &
1311 &
1289 &
355 &
2257 &
1848 &
2259 &
845 \\

\hskip -1mm
\includegraphics[width=0.08\columnwidth]{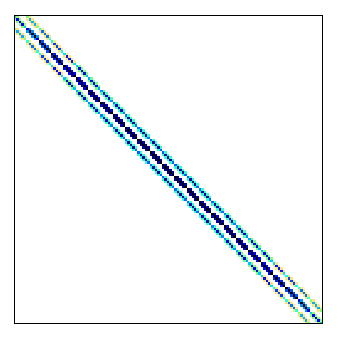} &
\includegraphics[width=0.08\columnwidth]{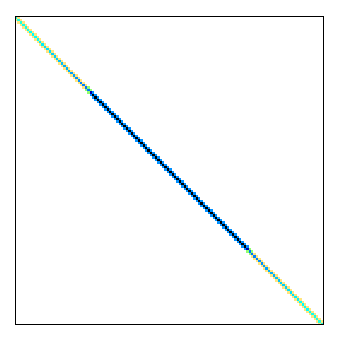} &
\includegraphics[width=0.08\columnwidth]{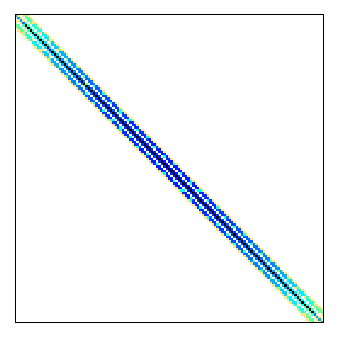} &
\includegraphics[width=0.08\columnwidth]{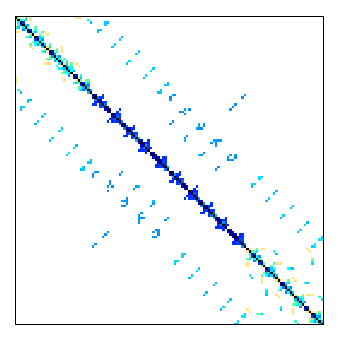} &
\includegraphics[width=0.08\columnwidth]{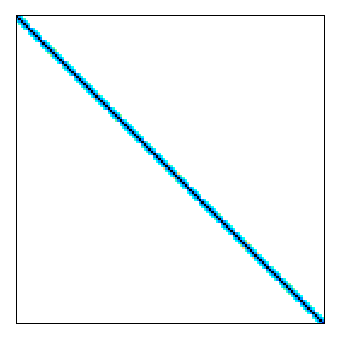} &
\includegraphics[width=0.08\columnwidth]{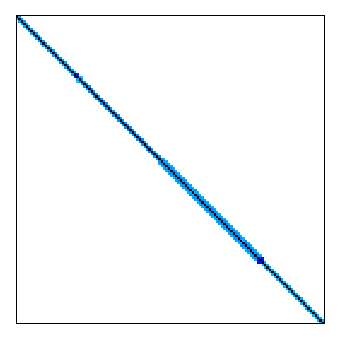} &

\includegraphics[width=0.08\columnwidth]{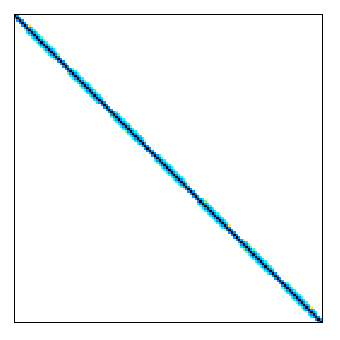} &
\includegraphics[width=0.08\columnwidth]{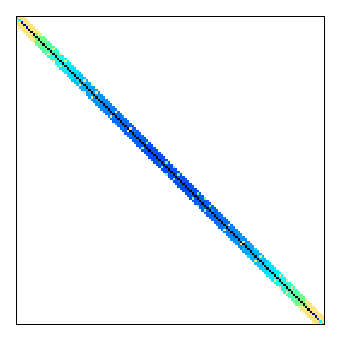} &
\includegraphics[width=0.08\columnwidth]{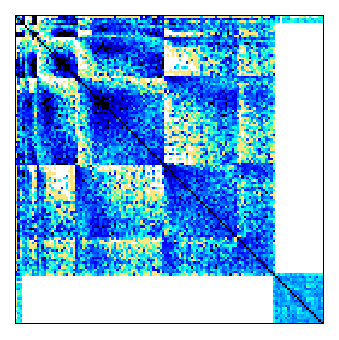} &
\includegraphics[width=0.08\columnwidth]{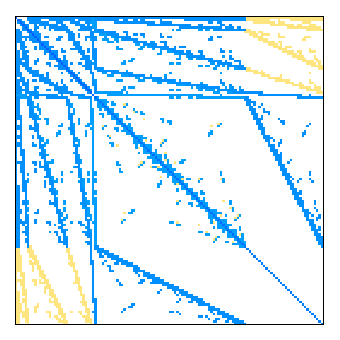} &
\includegraphics[width=0.08\columnwidth]{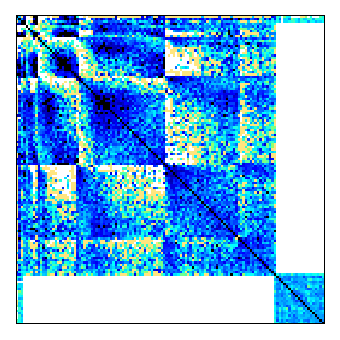} &
\includegraphics[width=0.08\columnwidth]{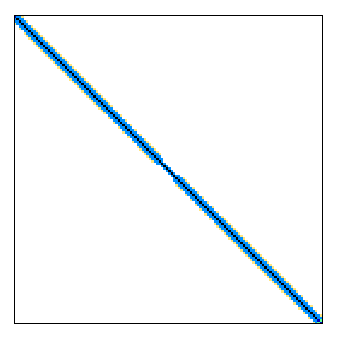} 
\hskip -2mm
  \end{tabular}
} \\
\hline
  \end{tabular}
  \label{table:matrices}
\vspace{-6pt}
\end{table}

\begin{figure*}[tb]
\centering
\includegraphics[width=1.8\columnwidth]{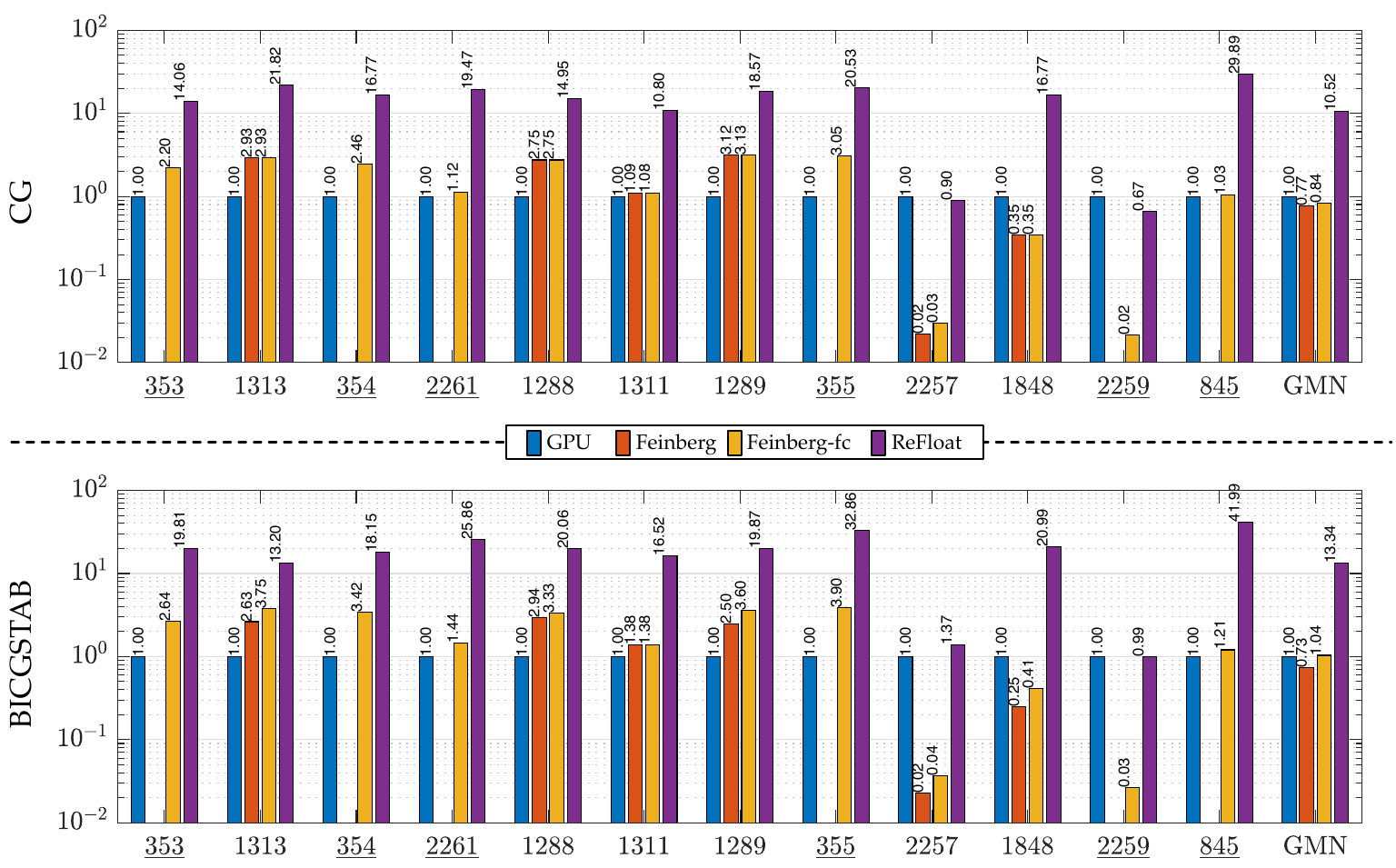}
\vspace{-6pt}
\caption{The performance of GPU, Feinberg~\cite{feinberg2018enabling}, Feinberg-fc and \rfloat for CG and BiCGSTAB solvers.}
\label{figure:cgbicgperf}
\vspace{-6pt}
\end{figure*}

We list the configurations for the baseline
GPU platform, the state-of-the-art ReRAM accelerator~\cite{feinberg2018enabling} for
scientific computing (Feinberg) and our \rfloat in Table \ref{table:platform}.
We use an NVIDIA
Tesla V100 GPU, which has 5120 Cuda 
cores and a 32GB HBM2 memory. We use CUDA version
11.7 and cuSPARSE routines in the iterative
solvers for the processing
on sparse matrices. We measure the running 
time for the solvers on the GPU.
For the two ReRAM accelerators, 
i.e. Feinberg~\cite{feinberg2018enabling} and \rfloat,
we use the parameters in 
Table \ref{table:platform} for simulation. 
Both the two ReRAM accelerators have 128 Banks
and the crossbar size is $128\times128$. 
In Feinberg~\cite{feinberg2018enabling}, we configure 64 clusters for each bank,
which is slightly larger than that (56) in Feinberg~\cite{feinberg2018enabling}. There
are 128 crossbars in each cluster. The precision
in Feinberg~\cite{feinberg2018enabling} is double floating-point. In \rfloat,
we configure 128 banks, 128 subbanks per bank, and 
64 crossbars per subbank. The precision in \rfloat
is \texttt{\textbf{refloat}} with a default setting
that $e=3$, $f=3$, $e_v=3$ and $f_v=8$.
For the two ReRAM 
accelerators, the equivalent computing ReRAM is
17.1Gb.
The ADC and ReRAM cells for the two accelerators
are of the same configuration. 
We use a 1.5GS/s 10-bit 
pipelined SAR ADC \cite{kull201728} for conversion.
The DAC is 1-bit, which is implemented by wordline
activation. We use 1-bit SLC \cite{niu2013design}
and the write latency is 50.88ns.
The computing latency for one crossbar,
including the ADC conversion, is 107ns~\cite{feinberg2018enabling}.

Table \ref{table:matrices} lists
the matrices used in the evaluation.
We evaluate on 12 solvable matrices from the 
SuiteSparse Matrix Collection (formerly UF Sparse Matrix Collection)
\cite{davis2011university}. The matrices' size 
(number of rows) ranges from
4,875 to 204,316 and the Number of Non-Zero 
entries (NNZ) of the matrices ranges from
105,339 for 1,660,579. NNZ/Row is a metric
for sparsity. A smaller NNZ/Row indicates
a sparser matrix.
NNZ/Row ranges from 4.0 to 27.7.
The condition number $\kappa$ ranges widely from 3.6 to 5.74e+5.
We also visualize the matrices in 
Table \ref{table:matrices}.
We apply the iterative solvers CG and BiCGSTAB
on the matrices. The convergence criteria
for the solvers is that the L-2 norm of the
residual vector (we use the term ``residual'' denoted by $R^2$ for simplicity to call 
the L-2 norm of the residual vector in this section) is less than $10^{-8}$.

\subsection{Performance}
\label{sec:perf_cg_bicg}

\begin{figure*}[t]
\centering
\includegraphics[width=1.8\columnwidth]{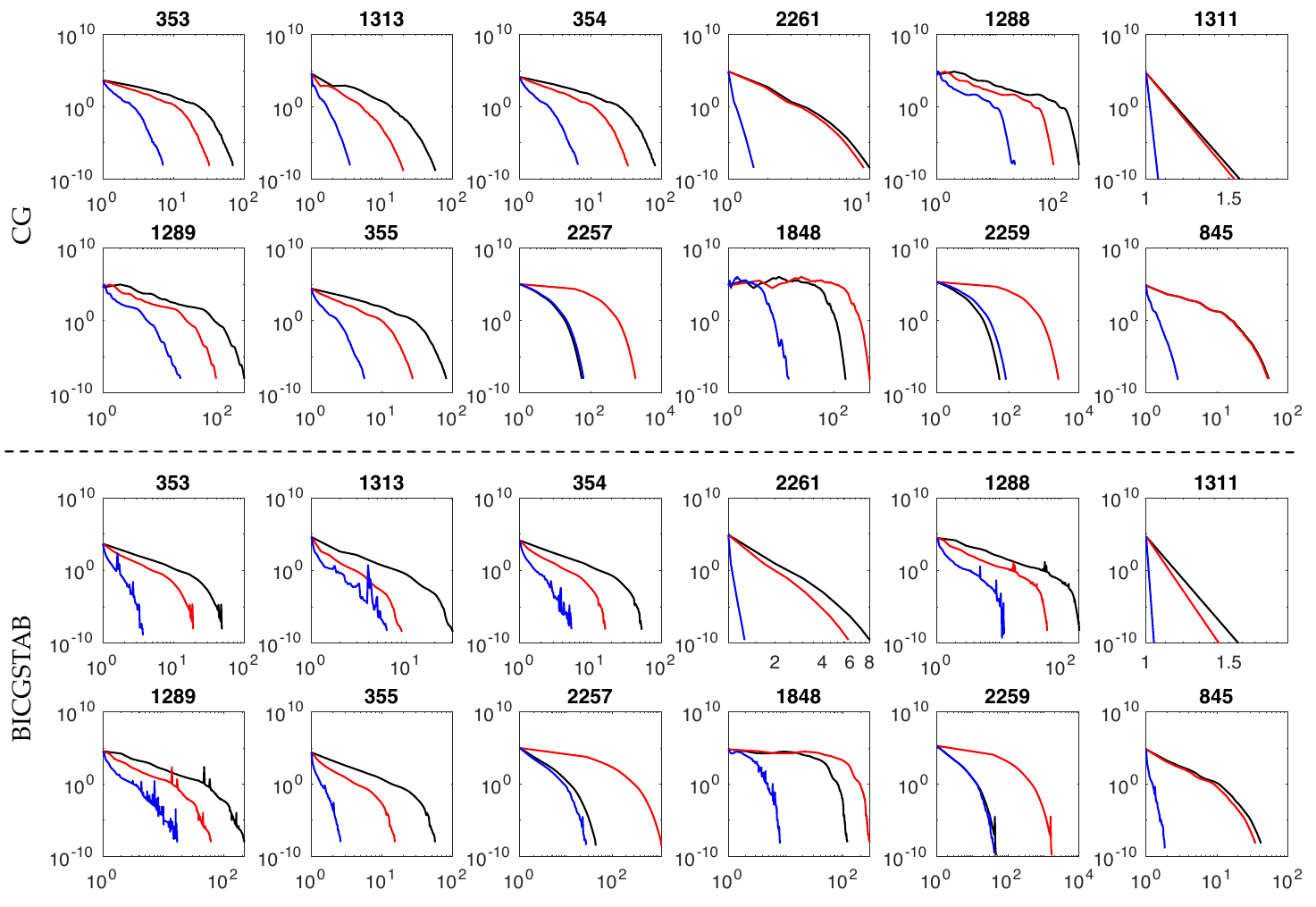}
\vspace{-6pt}
\caption{Convergence traces of CG and BiCGSTAB solvers of GPU (black line), Feinberg~\cite{feinberg2018enabling}-fc (red line) and \rfloat (blue line). The Y axis is the residual and the X axis is the
normalized (to GPU) iteration number.}
\label{figure:cgbicgcurve}
\vspace{-6pt}
\end{figure*}

We show the performance of the GPU, a state-of-the-art ReRAM accelerator Feinberg~\cite{feinberg2018enabling} and \rfloat for CG and BiCGSTAB
solvers in Figure~\ref{figure:cgbicgperf}. 
We evaluate the processing time $t$
for the iterative solver
to satisfy that the residual is less than $10^{-8}$.
The performance $p$ is defined
as $p=t_{\text{GPU}}/t_x$, $x=$ Feinberg~\cite{feinberg2018enabling},  Feinberg-fc or \rfloat.
For Feinberg~\cite{feinberg2018enabling},
we evaluate both function (convergence) 
and hardware performance.
Note that as we discussed in Sec.~\ref{sec:nc_feinberg},
the vector issue in~\cite{feinberg2018enabling} 
may lead to non-convergence on most matrices.
Feinberg-fc is a strong baseline
where we assume the function is 
correctly the same
as that of the default \texttt{\textbf{double}}. Specifically,
we assume Feinberg-fc 
converges and takes the
same iteration number
to convergence as that
in \texttt{\textbf{double}}
and evaluate the hardware performance of Feinberg-fc.

\noindent
\textbf{CG solver.}
Overall, the geometric-mean(GMN) performance of 
Feinberg~\cite{feinberg2018enabling}-fc 
and \rfloat
are $0.8362\times$ and $12.59\times$(up to $29.89\times$) respectively.
GPU and \rfloat converge on all matrices
while Feinberg~\cite{feinberg2018enabling} does not converge on 6 out of 12 matrices and the
IDs of not converged matrices are 
{\tt 353}, {\tt 354}, {\tt 2261}, {\tt 355}, {\tt 2259}, and {\tt 845}.
The GMN of \rfloat compared to Feinberg~\cite{feinberg2018enabling} on the
six converged matrices is $12.94\times$.
For most of the matrices,
\rfloat performs better than the
baseline GPU. For matrix {\tt 2257}, {\tt 1848} and
{\tt 2259}, the performance of \rfloat is
$0.8973\times$, $16.77\times$ and $0.6660\times$ respectively.
However,
the performance of Feinberg~\cite{feinberg2018enabling} is even lower, and
it is 2.21E-2$\times$, 3.48E-1$\times$ and NC respectively.
The slow down is because the required number of
clusters for SpMV is larger than the number
available on the accelerators.
If the number of clusters for SpMV on one matrix
is fewer than the available clusters on an accelerator,
the deployed clusters will be only invoked once
to perform the SpMV. But, if the
number of clusters for SpMV on one matrix
is larger than the available clusters on an accelerator,
(1) cell writing for mapping new matrix blocks to clusters and
(2) cluster invoking to perform part of SpMV will
happen multiple times, thus more time is consumed
for one SpMV on the whole matrix.
In Feinberg~\cite{feinberg2018enabling}, with
the default floating-point mapping, i.e., 118 crossbars for
a cluster, 
there are only 2221 clusters available.
However, to perform one SpMV on the whole matrix,
209263, 15797, and 381321 clusters are required respectively
for matrix {\tt 2257}, {\tt 1848}, and {\tt 2259}.
The required cluster number for the two matrix is far
larger than the available number in Feinberg~\cite{feinberg2018enabling}, resulting in
cell writing and cluster invoking 103, 8, and 187
times respectively for the three matrices.
So the performance
of Feinberg~\cite{feinberg2018enabling} is lower than the baseline GPU on the two matrices.
In \rfloat, 
to perform one SpMV on the whole matrix,
the same numbers as that in Feinberg~\cite{feinberg2018enabling}
of clusters are required for
 matrix {\tt 2257} and matrix {\tt 2259}.
We configure $e=3$, $f=3$ for \rfloat,
so the available clusters for matrix {\tt 2257}
and matrix {\tt 2259} are 21845. The 
cell writing and cluster invoking times for 
matrix {\tt 2257}
and matrix {\tt 2259} are 10 and 18 respectively,
which are less than the 
cell writing and cluster invoking times in Feinberg~\cite{feinberg2018enabling}.
\begin{table}[tb]
  \centering
  \caption{Absolute iteration number to reaching convergence.}
  \vspace{-6pt}
  \begin{tabular}{p{5mm}p{8mm}p{8mm}p{5mm}p{8mm}p{8mm}p{5mm}}
    \hline
    \multirow{2}{*}{\textbf{ID}} & \multicolumn{3}{c}{\textbf{CG}}&  \multicolumn{3}{c}{\textbf{BiCGSTAB}}\\
    & \texttt{\textbf{double}}& \texttt{\textbf{refloat}}& \textbf{+/-} & \texttt{\textbf{double}}& \texttt{\textbf{refloat}} & \textbf{+/-}\\
    \hline
{\tt 353}  & 68         & 85  & +17 & 49 & 51 & +2\\
{\tt 1313} & 52         & 55  & +3  & 34 & 69 & +35\\
{\tt 354}  & 81         & 95  & +14 & 58 & 79 & +21\\
{\tt 2261} & 11         & 11  & 0   & 7  & 7  & 0\\
{\tt 1288} & 262        & 305 & +43 & 195 & 205 & +10\\
{\tt 1311} & 1          & 1   & 0   & 1  & 1  & 0\\
{\tt 1289} & 294        & 401 & +107& 211 & 317 & +106\\
{\tt 355}  & 80         & 95  & +15 & 59 & 52 & -7\\
{\tt 2257} & 55         & 56  & +1  & 43  & 36 & -7\\
{\tt 1848} & 162        & 214 & +52 & 118 & 145 & +27\\
{\tt 2259} & 57         & 58  & +1  & 45 & 36 & -9\\
{\tt 845}  & 53         & 54  & +1  & 41 & 35 & -6\\
    \hline
  \end{tabular}
  
  \label{tab:converge}
\vspace{-6pt}
\end{table}

Another reason leading to higher performance of \rfloat
compared with Feinberg~\cite{feinberg2018enabling} is that fewer cycles are consumed within
a cluster. In Feinberg~\cite{feinberg2018enabling}, 233 cycles are consumed for the
multiplication even with the assumption that 6 bits are
enough for the exponent~\cite{feinberg2018enabling}. 
In \rfloat, 28 cycles are consumed for the
multiplication. 
Notice that with a fewer number of exponent bits and
fraction bits, we can get (a) a fewer number of clusters
required for a whole matrix, (b) a fewer number of cycles
consumed for one matrix block floating-point multiplication
within a cluster. The two effects (a) and (b) can lead to
higher performance, but we also have a third effect (c)
larger number of iterations to reaching convergence, which leads
to lower performance. However, effects (a) and (b)
is stronger than effect (c),
so the performance of \rfloat is higher. The number
of iterations for the evaluated matrices to reach convergence
is listed in Table \ref{tab:converge}.

\noindent
\textbf{BiCGSTAB solver.}
The geometric-mean(GMN) performance of 
Feinberg~\cite{feinberg2018enabling}-fc 
and \rfloat
are $1.036\times$ and $13.34\times$ (up to $41.99\times$) respectively.
The GPU and \rfloat converge on all matrices
while Feinberg~\cite{feinberg2018enabling} does not converge on 6 out of 12 matrices and the
IDs of not converged matrices are 
{\tt 353}, {\tt 354}, {\tt 2261}, {\tt 355}, 
{\tt 2259}, and {\tt 845}.
The GMN of \rfloat compared to Feinberg~\cite{feinberg2018enabling} on the
four converged matrices is $15.98\times$.
The trend for the three platforms on the evaluated matrices
are similar to that for CG solver. In each iteration, for
CiCGSTAB solver, there are two SpMV on the whole matrix, while
for CG solver, there is one SpMV on the whole matrix.
From Table \ref{tab:converge} we can see, 
the difference of (\textbf{+/-})
number of iterations to get converge in BiCGSTAB
solver is smaller than the gap in CG solver for most matrices.
For matrix {\tt 355}, {\tt 2257}, {\tt 2259} and {\tt 845}, 
the difference is negative, which means it takes fewer iterations in
\texttt{\textbf{refloat}} compared with that in \texttt{\textbf{double}}.

\subsection{Accuracy}

\begin{table}[tb]
  \centering
  \caption{Bit number for exponent and fraction of matrix block and vector segment in \rfloat.}
  \vspace{-6pt}
  \begin{tabular}{llllllll}
    \hline
     \multicolumn{4}{c}{\textbf{CG}}&  \multicolumn{4}{c}{\textbf{BiCGSTAB}}\\
    $e$ & $f$ & $e_v$ & $f_v$ & 
    $e$ & $f$ & $e_v$ & $f_v$ \\
    \hline
3  & 3  & 3  & 8  & 3  & 3  & 3  & 8 \\
    \hline
  \end{tabular}
  \label{tab:bitnum}
\vspace{-6pt}
\end{table}

We show the convergence traces (the residual
over each iteration) of GPU, Feinberg-fc, and \rfloat
for CG and BiCGSTAB solvers in Figure~\ref{figure:cgbicgcurve}.
The iteration number is normalized by the consumed time for the GPU baseline. Table \ref{tab:bitnum} lists the configurations of bit number for matrix block and
vector segment in \texttt{\textbf{refloat}}
for all matrices except
{\tt 1288} and {\tt 1828}.
For {\tt 1288} and {\tt 1828},
the only difference is the $f_v=16$.
The absolute (non-normalized)
iteration number to reach convergence 
is listed in Table \ref{tab:converge}.

For CG solver, from Table \ref{tab:converge} we can see,
\texttt{\textbf{refloat}} leads to more number of
iterations to get converged when we do not consider the
time consumption for each iteration. From 
Figure~\ref{figure:cgbicgcurve} we can see, with the low bit
representation, the residual curves are almost
the same trend as the residual curves of GPU and Feinberg-fc in default
\texttt{\textbf{double}}. Most importantly,
all the traces in \texttt{\textbf{refloat}} 
get converged faster than the
traces of GPU and Feinberg-fc.
For matrix {\tt 1288} and matrix {\tt 1848}, the bit number
for fraction of vector segment is 16 because the default 8 
leads to non-convergence.
For BiCGSTAB solver, from Table \ref{tab:converge} we can see,
while \texttt{\textbf{refloat}} leads to more number of
iterations to reaching convergence for 5 matrices, the 
number of iterations to reaching convergence for 4
matrices are even fewer than those in \texttt{\textbf{double}}.
We infer that is because lower bit representation helps
to enlarge the changes in the correction term, thus leads
to fewer iterations. 
We also notice there are spikes in the residual curves in
\texttt{\textbf{refloat}} more frequently than spikes in
\texttt{\textbf{double}}, but they finally reach convergence.

\subsection{Robustness to Noise}
\label{sec:tobustness}

\begin{figure}[tb]
\centering
\includegraphics[width=0.85\columnwidth]{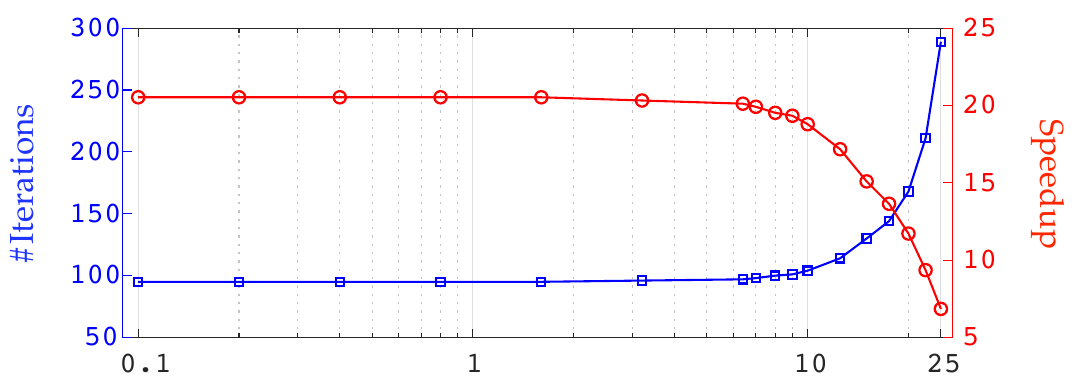}
\vspace{-6pt}
\caption{The iteration number and speedup of \rfloat on {\tt crystm03} v.s. noise.}
\label{figure:noise_perf}
\vspace{-6pt}
\end{figure}

\begin{table}[t]
\vspace{0pt}
  \centering
  \caption{Memory overhead of \rfloat v.s. Feinberg~\cite{feinberg2018enabling}.}
  \vspace{-6pt}
  \small
\begin{tabular}{rllllllllllll}

\hline

\textbf{ID} 
& {\tt 353} & {\tt 1313} & {\tt 354}  
& {\tt 2261} & {\tt 1288} & {\tt 1311} 
\\

& 0.173 & 0.176 &  0.173 
& 0.176 & 0.173 &  0.174 
\\

\hline

\textbf{ID} 
& {\tt 1289} & {\tt 355}  & {\tt 2257} 
& {\tt 1848} & {\tt 2259} & {\tt 845}\\

&  0.173 &  0.173  &  0.312
&  0.179 &  0.300 &  0.173\\

\hline

\end{tabular}
   
\label{tab:memoryoverhead}
\vspace{-6pt}
\end{table}

To study the robustness to noise of \rfloat, we disable the
error correction.
We model the random telegraph noise (RTN) \cite{choi2014random}
which is widely adopted in ReRAM accelerator noise modeling  
\cite{hu2016dot,feinberg2018enabling,agarwal2016resistive}.
We use {\tt crystm03} with CG solver
for a case study and show
the speedup (compared to GPU) and iteration number
v.s. noise deviation $\sigma$ from $0.1\%$ to $25\%$
in Figure~\ref{figure:noise_perf}.
Within $10\%$ noise, the speedup degrades very little
and at $25\%$ noise, \rfloat still maintains a $6.85\times$ speedup. As we discussed before,
the iterative solvers naturally tolerate noise
and deviation.

\subsection{Memory Overhead}
In Table \ref{tab:memoryoverhead},
we compare the memory overhead for the matrix in
\texttt{\textbf{refloat}} normalized to that in \texttt{\textbf{double}}
(used in Feinberg~\cite{feinberg2018enabling}).
On average, \texttt{\textbf{refloat}} consumes $0.192\times$
memory compared with \texttt{\textbf{double}}.
For matrices except {\tt 2257} and {\tt 2259},
\texttt{\textbf{refloat}} consumes less than $0.2\times$
memory compared with \texttt{\textbf{double}}.
For matrix {\tt 2257} and matrix {\tt 2259},
the average density within a matrix is relatively lower,
thus more memory is consumed for
the matrix block index and the exponent base.

\section{Related Works}
\label{sec:relatedworks}

\noindent
\textbf{ReRAM-based accelerators.}
In recent years, the architecture design of ReRAM-based accelerators
have been developed for various applications, including
deep learning \cite{bojnordi2016memristive,chi2016prime,shafiee2016isaac, song2017pipelayer,imani2019floatpim,yang2019sparse,ji2019fpsa,ankit2019puma,imani2020deep,feinberg2018making},
graph processing \cite{song2018graphr,challapalle2020gaas} and scientific computing \cite{feinberg2018enabling}.
The noise and reliability issues in ReRAM-based computing 
are significantly alleviated by coding techniques and architectural optimizations 
\cite{feinberg2018making,liu2019fault,wen2020accelerating,wen2019renew,wen2018wear}.
ReRAM-based accelerators are demonstrated on 
silicon by \cite{wu2018brain,chen201865nm,xue201924,pang201925,xue202015,wan202033,liu202033}.
Most ReRAM-based accelerators
are designed for fixed-point processing, especially for deep learning.
Besides \cite{feinberg2018enabling},
\cite{feinberg2021analog} applied preconditioner
and
FloatPIM \cite{imani2019floatpim} accelerated floating-point multiplication in ReRAM, but
FloatPIM is designed for deep learning 
in full-precision floating point.

\noindent
\textbf{Scientific computing acceleration.}
Computing routines on general-purpose platforms CPUs and GPUs
have been developed for scientific computing, such as
CuSPARSE \cite{naumov2010cusparse}, 
MKL \cite{wang2014intel},
and LAPACK \cite{anderson1999lapack}.
Architectural and architecture-related optimizations 
\cite{feng2021egemm,zhang2017understanding,liu2018towards,dakkak2019accelerating,jia2018dissecting,lai2013performance,shen2018cudaadvisor}
on CPUs/GPUs are explored
for accelerating scientific computing. \cite{dong2020smart,dong2019adaptive,dong2023auto} leveraged machine learning for the acceleration of scientific computing and \cite{song2022sextans, song2022serpens, song2023callipepla} accelerated sparse linear algebra and solvers on FPGAs.
Scientific computing is a major application
in high performance computing 
and heavily 
relies on
general-purpose platforms, but it is 
a new application domain for emerging
PIM architectures 
and it is challenging
because of high cost and low performance
of floating-point processing.

\noindent
\textbf{Data format.}
Data formats for efficient computing
are explored for CPUs/GPUs
\cite{liu2015csr5,li2018hicoo,smith2015tensor,bader2008efficient,jeon2015haten2,liu2017unified}.
Format and architecture co-optimization
includes \cite{han2016eie,srivastava2020tensaurus,fowers2014high} on CMOS platforms
but they are not for emerging PIM architectures
and not for scientific computing.
Data compression are explored on
DRAM systems~\cite{pekhimenko2012base, pekhimenko2013linearly,lee2015warped}.

\section{Conclusion}
\label{sec:conclusion}
ReRAM has been proved
promising for accelerating 
fixed-point applications
such as machine learning,
while scientific computing is an application domain that
requires floating-point processing.
The main challenge for efficiently accelerating
scientific computing in ReRAM is how to support
low-cost floating-point SpMV in ReRAM.
In this work, 
we address this challenge by proposing \rfloat, 
a data format, and a
supporting accelerator architecture.
\rfloat is tailored for processing on ReRAM crossbars.
The number of effective bits is significantly reduced
to reduce the crossbar cost and cycle cost for
the floating-point multiplication on a matrix block.
The evaluation results across a variety of benchmarks reveal 
that the \rfloat accelerator
 delivers 
a speedup
of 
$5.02\times$ to $84.28\times$
compared with a state-of-the-art
ReRAM-based accelerator \cite{feinberg2018enabling} for scientific computing even with the assumption that the accelerator~\cite{feinberg2018enabling} functions the same as FP64 solvers.
We released the source code at \url{https://github.com/linghaosong/ReFloat}.

\bibliographystyle{IEEEtranS}
\bibliography{main.bib}

\end{document}